# Variations of TEC over Iberian Peninsula in 2015 due to geomagnetic storms and solar flares

### A. L. Morozova, T. V. Barlyaeva, and T. Barata

Univ Coimbra, Center for Earth and Space Research of the University of Coimbra, Almas de Freire, Sta. Clara, 3040-004, Coimbra, Portugal

Corresponding author: Anna L. Morozova (annamorozovauc@gmail.com)

**Key Points:**

- Variations of TEC over the Iberian Peninsula associated with geomagnetic storm and solar UV flux variations are analyzed using PCA

- A combination of PCA and multiple regression models with space weather parameters as regressors is well suited for TEC reconstruction

- PCA-MRM reconstruction has a good potential for TEC forecasting

**Abstract**

The total ionospheric content (TEC) over the Iberian Peninsula was studied using data from two locations obtained both by GNSS receivers and an ionosonde. The principal component analysis applied to the TEC data allowed us to extract two main modes. Each mode is characterized by daily TEC variation of a certain type (PC) and its amplitude for each of the studied day (given by the empirical orthogonal functions, EOF).

The variations of these modes as well as the original TEC data were studied in relations to four strongest geomagnetic storms of 2015 and three geomagnetic disturbances of lower amplitude observed during the same months. EOFs were found to correlate well with space weather parameters characterizing solar UV and XR fluxes, number of the solar flares, parameters of the solar wind and geomagnetic indices.

Multiple regression models were constructed to fit EOFs using combinations of the space weather parameters with a lag from 0 to 2 days. Combining the regression models of EOFs with the corresponding PCs we reconstructed TEC variations as a function of space weather parameters observed in previous days. The possibility to use such reconstructions for the TEC forecasting was also studied.

**1 Introduction**

Modern society, industry and science use GNSS-based technologies more and more often: operation of unmanned air-born, floating and land vehicles, in-route and assisted landing procedures for commercial aviation, rescue operations, etc. The quality of the GNSS signal and therefore the reliability of the GNSS-based technological solutions depend on the conditions in the upper part of the Earth atmosphere – the partially ionized layer called ionosphere.

The main sources for the data on the ionospheric conditions are ionosondes installed at specific locations  and networks of GNSS receivers. The ionosondes provide information about the altitude and peak electron density for different ionospheric layers (D, E, F1 and F2) allowing, consequently, to estimate a so called ionospheric total electron content (iTEC) without plasmaspheric contribution. However, the number of ionosondes is limited (e.g., the Ebro Observatory, Spain, owns the only regularly working ionosonde on the Iberian Peninsula); they provide information for a certain area and with limited time resolution. As an alternative, the territories of heavily populated areas like the continental Europe are well covered by networks of GNSS receivers. These networks allow to monitor ionospheric conditions with high precision and to develop empirical models to predict ionospheric response to different external forcings, e.g., solar flares and geomagnetic storms.

One of the widely used parameters characterizing ionospheric conditions that can be obtained from the GNSS data is the total electron content (TEC). TEC data are often used [*Astafyeva et al.,* 2017; *Goncharenko et al.*, 2013; *Li et al.,* 2019] to analyze and model ionospheric disturbances caused by different events. In general, TEC response to space weather events (like flares or geomagnetic storms) consists of either changes of the amplitude and the shape of the regular daily TEC variation observed world-wide or a phenomenon called travelling ionospheric disturbances (TID) which are observed on smaller spatio-temporal scales. In particular, it was shown that at middle latitudes ionospheric conditions (TEC, TIDs and scintillation events) respond significantly to strong geomagnetic storms of recent years [*Aa et al.,* 2018; *Astafyeva et al.,* 2015; *Shim et al.,* 2018]. These variations can be a reason for the GNSS signal degradation and a decrease of the precision of the positioning in the affected area.

The results of the data analysis can be used to develop regional empirical models allowing prediction of ionospheric response to different forcings, e.g., geomagnetic storms. In particular, the method called principal component analysis (PCA) showed good potential for the analysis and modeling of ionospheric regular variations and disturbances [*Chen et al.,* 2015; *Li et al.,* 2019; *Morozova et al.,* 2019].The advantage of the regional models [e.g., *Hu and Zhang,* 2018; *Mukhtarov et al.* 2017; *Petry et al.,* 2014; *Tebabal et al.,* 2019; *Tsagouri, et al.,* 2018] over the global ones is in their better reflection of specific local behavior of the ionosphere.

Large national and international research centers and companies can afford development and operation of complex physics-based models for the upper atmosphere and ionosphere. However, for smaller scale private enterprises such models are too expensive and demanding in human resources and computational costs. Thus, there is a demand from small enterprises providing services for local GNSS users for simplistic models forecasting ionospheric conditions for the following 1-2 days. Such models have to be cheap in maintenance, use reasonable amount of proprietary products and provide reliable forecasts. One of the optimal candidates is the class of statistical models based on the regression of the ionospheric parameters on a set of parameters known to be forcings or precursors of ionospheric disturbances, such as the solar UV and XR irradiance, daily number of solar flares, parameters of the solar wind and geomagnetic indices.

In this paper we present results of the exploratory analysis of the TEC data obtained for a mid-latitudinal region both from the ionosonde measurements and from the data of GNSS receivers. The main goal of this study is to characterize the response of the ionosphere to such events like geomagnetic disturbances, solar flares and variations of the solar UV and XR fluxes, test regression models based on different sets of regressors (solar (fluxes and flares), solar wind and geomagnetic parameters) and with different time lags, and explore their forecasting abilities.

The paper is organized as follows. Section 1 gives an introduction. Sections 2 and 3 describe, respectively, the data in use and the methods. Section 4 gives the comparison of different TEC data series and their PCA reconstructions. Section 5 presents analysis of the TEC variations that observed during March, June, October and December of 2015 in relations to the geomagnetic disturbances, solar flares and solar UV and XR fluxes variations occurred during these month. Section 6 presents the results of the multiple regression analysis of TEC variations on the solar and geomagnetic parameters. The final conclusions are given in Section 7.

## 2 Data

In this work we used following data sets from different ground and space missions.

### 2.1 Total electron content

The TEC series are obtained for two locations on the Iberian Peninsula (IP), on its west and east coasts, respectively, Lisbon (Portugal) and Ebro (Spain),. Three sources for the vertical TEC data were used in this work.

First data set is the GNSS TEC data from the Royal Observatory of Belgium (ROB) data base available as vertical TEC maps in the IONEX format on a grid of 0.5° x 0.5° with 15 min time resolution [see also *Bergeot et al*., 2014]. Accordingly to the information from the ROB web site, the vertical TEC is estimated in near real-time from the GPS data of the EUREF

Permanent Network (EPN). The vertical TEC maps are produced from the slant TEC of each satellite-receiver pair as projections in the vertical TEC at the ionospheric piercing points (ionospheric shell at 450 km) and interpolated to a grid using splines. Here the TEC data for the grid points most close to Lisbon (39°N, 9°W) and Ebro (41°N, 0.5°E) were used, hereafter *ROB-LIS* and *ROB-EBR*, respectively.

Another series, hereafter *SCI-LIS*, is the GNSS TEC measured by the SCINDA receiver been active from the November 2014 to the July 2019 in the Lisbon airport (38.8°N, 9.1°W) in the frame of the ESA Small ARTES Apps project SWAIR (Space Weather and GNSS monitoring services for Air Navigation). The installed equipment is a NovAtel EURO4 receiver with a JAVAD Choke-Ring antenna. The data originally of 1 min time resolution were averaged to obtain the 1 h time resolution. The details of the SCINDA software functionalities can be found in [*Carrano et al.*, 2009a, b; *Carrano and Groves*, 2007, 2009]. Since the data are provided in archived format [*Barlyaeva et al.*, 2020a, b] and contain some peculiarities (i.e. missed or erroneous blocks) to be taken into account and preprocessed, a software named "SCINDA-Iono" toolbox for MATLAB [*Barlyaeva et al.*, 2020b] was developed by our group. The calibration procedure was not performed during the installation of this receiver. To validate the non-calibrated *SCI-LIS* series we present in Section 4 a comparison of *SCI-LIS* with other TEC series.

Third data source is the iTEC (*IONO-EBR*) provided by the *Ebro Observatory*, Spain (40.8°N, 0.5°E, 50 m asl). The instrument currently installed at the Ebro Observatory is the DPS-4D ionospheric sounder and the measured parameter is the critical frequencies of the ionospheric layers. The altitude profiles of electron density are calculated from the ionograms, and the integration of these electron profiles up to 1000 km height gives the values of iTEC [*Huang* and *Reinisch*, 2001]. The *IONO-EBR* data are of 1 h time resolution.

The 1 h series of TEC for the four months of 2015 (March, June, October, and December) are shown in the Supporting Information (SI) Fig. S1. The time is UT (for Lisbon UT = LT, for Ebro UT = LT-1); since the longitudinal distance between Lisbon and Ebro is ~9.5° solar time difference is ~40 min.

### 2.2 Space weather data

Five parameters were used to characterize geomagnetic field variations. The global *Dst* index have 1 h time resolution. The global *Kp* and *ap* indices and the local K-index ($K_{COI}$, calculated from the horizontal component of the geomagnetic field measured at the *Coimbra Magnetic Observatory,* COI, Coimbra, Portugal, 40.2°N, 8.4°W, 99 m asl) originally have 3 h time resolution. The fifth index is the auroral electrojet (*AE*) index characterizing the auroral activity in the polar regions. Weak and strong geomagnetic disturbances (with *Dst* ≤ -50 nT) were analyzed separately with a threshold between them set at –100 nT.

The data on the solar wind properties were obtained from the OMNI data base. In this study we used following parameters with the original 1 h time resolution: interplanetary magnetic field (IMF) components as scalar *B, Bx,* GSM *By* and *Bz* (in nT), solar wind flow speed (*v* in km/sec), proton density (*n* in n/cm$^3$) and flow pressure (*p* in nPa).

To parameterize the variations of the solar UV radiation we used two proxies. The first one is the *Mg II* composite series [*Snow et al*., 2014], a proxy for the spectral solar irradiance variability in the spectral range from UV to EUV based on the measurements of the emission core of the Mg II doublet (280 nm). The second proxy is the *F10.7* index from the OMNI data base. Both *Mg II* and *F10.7* series have 1 d time resolution.

As a proxy for the variations of the solar XR flux we used the data measured by the Solar EUV Experiment (SEE) for the NASA TIMED mission at the wavelength 0.5 nm with time resolution of 1 d (*XR*). Please note that for the plots the UV and XR irradiance proxies were scaled individually for better visualization.

The information about the solar flares observed during the analyzed time interval was obtained through the NOAA National Geophysical Data Center (NGDC). The daily numbers of solar flares of classes C, M and X separately as well as the total daily number of such flares (N) were calculated. Only flares occurred during the local day time were considered.

The variations of the geomagnetic indices, the solar UV and XR fluxes, the number of solar flares as well as parameters of the solar wind and the interplanetary magnetic field for 2015 are presented in Fig. S2 in SI (daily means).

## 3 Methods

All series used in the presented work were averaged or interpolated to obtain series with 1 h and 1 d time resolution. A number of small gaps were linearly interpolated.

The analysis of the TEC series obtained from different sources was performed using the principal component analysis (PCA). The input data set is used to construct a covariance matrix and calculate corresponding eigenvalues and eigenvectors. The eigenvectors are used to calculate principal components (PC) and empirical orthogonal functions (EOF). The combination of a PC and the corresponding EOF is called a "mode", and the eigenvalues allow to estimate the explained variances of the extracted modes. PCs are orthogonal and conventionally non-dimensional. The full descriptions of the method can be found in (e.g.) *Bjornsson and Venegas* [1997], *Hannachi et al.* [2007] and *Shlens* [2009].

Each of the TEC series obtained at different locations and by different instruments (*ROB-LIS, ROB-EBR, SCI-LIS* and *IONO-EBR*) with 1 h time resolution was analyzed separately during each month-long time interval. The PCA input matrices were constructed in a way that each column contains 24 observations (every 1 h) for a specific day. Thus, PCA allows us to obtain daily variations of different types as PCs and the amplitudes of those daily variations for each day of a month as corresponding EOFs. Thus, we need no models for the daily variation of TEC or its seasonal changes as function of zenith angle or DOY: the amplitude and the phase (daily maximum local time) of the daily TEC variation are automatically extracted from the observational data. Further, the 1st and 2nd PCA modes were used to reconstruct TEC variations.

Similarities between the variations of TEC and space weather parameters were analyzed using the correlation coefficients, *r*, that test linear relations between analyzed variables. The significance of the correlation coefficients was estimated using the Monte Carlo approach with artificial series constructed by the "phase randomization procedure" [*Ebisuzaki*, 1997]. The obtained statistical significance (*p value*) considers the probability of a random series to have the same or higher absolute value of *r* as in the case of a tested pair of the original series.

The linear multiple regression models (MRM) were constructed using TEC-related series (1 h and 1 d TEC, EOF1 and EOF2) as dependent variable, and solar (i.e., UV and XR fluxes and the flare numbers), solar wind and geomagnetic parameters as regressors. A "best subset" technique was chosen to ensure that only those regressors that are most influential for a particular TEC series were selected. The "best subset" was estimated using a so-called adjusted squared coefficient of multiple determination ($R_{adj}^2$). The value $R_{adj}^2 \cdot 100\%$ shows the percent of the variations of the dependent parameter explained by the MRM in question.

To estimate the quality of the MRMs we also used following statistics: $r$ – the correlation coefficient between the observed and modeled series; $RMSE/\sigma$ – root mean square error in units of the standard deviation ($\sigma$), see Eq. 1; the $\chi^2$ goodness-of-fit test (Eq. 2)

$$\frac{RMSE}{\sigma} = \frac{1}{\sigma}\sqrt{\frac{1}{N}\sum(TEC - TEC_{MRM})^2} \qquad (1)$$

$$\chi^2 = \sum \frac{(TEC - TEC_{MRM})^2}{TEC} \qquad (2)$$

where *TEC* is the observed and *TEC$_{MRM}$* is the modeled TEC series, and *N* is the length of the data series.

## 4 Comparison of TEC series from different locations and instruments and PCA reconstructions

In this work we used series of TEC obtained for two different locations in IP and by different instruments. Therefore, before further analysis we want to discuss similarities and differences of these series during the analyzed time interval.

The correlation analysis of the TEC series from different locations shows high correlation: mean value for the four analyzed months $r = 0.93$. There is also good agreement between the TEC values measured by the GNSS receivers and the ionosonde: mean $r = 0.95$. The *SCI-LIS* series is well correlated both with other GNSS-based TEC series (mean $r = 0.95$) and with iTEC (mean $r = 0.89$). The fact that *SCI-LIS* is not calibrated does not affect the results of our analysis since we used only methods insensitive to scaling and shifting of the series. Please note also that for the plots the *SCI-LIS* series was scaled to achieve better visualization (see Figs. S1, S3-S4). All correlation coefficients have *p values* $\leq 0.05$. Individual correlation coefficients can be found in Tab. S1 (SI).

There is a systematic difference between the ROB and ionosonde series: the GNSS-based TEC values are higher than the iTEC series by ~3-5 TECU, which, to our mind, reflect both the difference between the iTEC and TEC values [see discussion about plasmaspheric contribution in, e.g., *Belehaki et al.*, 2003, 2004; *McKinnell et al.*, 2007] and the fact that the ROB data are interpolation of the actual observations to a regular grid.

Since the TEC series are highly correlated, we calculated the mean TEC series for each month using the normalized (varying between –1 and +1) individual TEC series. The mean TEC series are in arbitral units (a.u.).

The individual TEC series for each month were submitted to PCA as described in Section 3. Here we will discuss only first two PCA modes, Mode 1 and Mode 2, respectfully. The mean PC1 and PC2 for both modes are shown in Fig. 1 and PCs for individual series are presented in Fig. S3-S4 (in SI). Corresponding mean EOF1 and EOF2 are shown in Figs. 2 and 3, respectively, and the individual EOFs are shown in Fig. S5-S6 (in SI).

Mode 1 explains 93-95%, 77-86%, 92-95%, and 87-94% of the TEC variations for March, June, October and December, respectively. As one can see in Fig. 1, PC of Mode 1 represents daily TEC variations caused by the regular changes of the insolation through a day. Daily minimum is observed during the local night and the daily maximum is in the afternoon hours. The differences between mean PCs for different months reflect seasonal changes of the ionospheric insolation and other climatologic conditions in the upper atmosphere.

Mode 2 of TEC series is defined as PC2/EOF2 for March and June for all TEC series and for October and December for the GNSS TEC series, whereas for the iTEC series this mode rather consists of PC3/EOF3. Mode 2 represents daily TEC variations with a relatively shallow minimum around the noon and a maximum in the late afternoon (19-21 h). Mode 2 explains 2.4-2.9%, 6.1-8.4%, 1.5-3.0%, and 2.5-3.7% of the TEC variations for March, June, October, and December, respectively. The variations associated with Mode 2 are relatively well correlated between the locations and instruments (mean r = 0.71, see also Tab. S1). Please note that for October and especially for December 2015 the iTEC Mode 2 shows lower correlation with other (GNSS) TEC series. The reason for this, probably, is that the variations associated with this mode for these months for the iTEC series are distributed between modes PC3/EOF3 (more) and PC2/EOF2 (less).

Overall, first two PCA modes together explain 90-95% of the variations of the original series. Therefore, we used these modes to make the PCA reconstructions of TEC. The mean TEC Mode 1 and Mode 2 series can be found in Figs. S7-S10.

## 5 TEC variations in 2015

In this work we considered variations of the ionospheric and space weather parameters during four months of 2015: March, June, October and December. Major geomagnetic storms with $Dst < -200$ nT occurred in March and June 2015 (strongest and second strongest geomagnetic storms of $24^{th}$ solar cycle). Two strong geomagnetic storms with $Dst \leq -125$ nT were observed in October and December of this year. Three of these four storms were associated with coronal mass ejections (CME). Also, in June, October and December weaker geomagnetic disturbances with $Dst < -50$ nT took place.

On March 11, 2015, an X2.1-class solar flare was observed between 16:11 UT and 16:29 UT. Even so that the X-class flares are expected to increase ionization in the ionospheric D, E, and F regions, this particular flare being of short durations (few minutes) did not have significant effect on TEC values [see *de Abreu et al.*, 2019, for the Brazilian sector].

During all four analyzed months there were series of solar flares of the C and M classes (see blue and red bar plots in Fig. S2, top left, in SI). On the whole, the increased number of flares resulted in the elevated level of the UV (*F10.7* and *Mg II* indices) and XR solar fluxes (see lines in Fig. S2, top left, in SI). Correlation coefficients between the series of daily means of *F10.7, Mg II* and *XR*, and the total daily number of the flares (all flares and C flares) are within 0.27-0.55, 0.23-0.46 and 0.22-0.71, respectfully, with higher and statistically significant (*p values* $\leq 0.05$) correlation coefficients for the *XR* proxy.

Also, on March 20 a partial solar eclipse with the maximal obscuration of 60% at 09:00 UT (IP area) took place resulting in the anomalously low amplitude of the TEC daily variations.

5.1 Geomagnetic storms

Geomagnetic storm on March 17-18, 2015 (GS1, see Tab. S2 in SI) is characterized by *Dst* < -220 nT and *Kp* = 8. This storm and its effect on the ionosphere is thoroughly analyzed by many authors [see, e.g., *Astafyeva et al.,* 2015; *Balasis et al.,* 2018; *Fagundes et al.,* 2016; *Liu et al.,* 2015; *Liu and Shen,* 2017; *Nava et al.,* 2016; *Paul et al.*, 2018; *Piersanti et al.,* 2017]. The storm was triggered by two flares of C type on March 14 and 15 and two halo CMEs arrived at 1 AU on March 17 [*Liu et al.,* 2015]. The recovery phase of the storm started on March 18 and lasted for several days; *Dst* remained below –50 nT until the end of March.

The Earth's ionosphere responded to this storm by an increase of TEC during the main phase ("positive ionospheric storm", see *Maruyama et al.* [2009]) on March 17 which was followed by a decrease of TEC ("negative ionospheric storm", *Maruyama et al.* [2009]) during March 18-19 [*Astafyeva et al.,* 2015; *Fagundes et al.,* 2016; *Liu and Shen,* 2017]. For Chinese middle latitudes the decrease of TEC relatively to the undisturbed level was ~60-70% [*Liu and Shen,* 2017]. In the European-African sector the positive storm on March 17 was observed only in the Northern Hemisphere [*Astafyeva et al.,* 2015]; and the daily TEC variations at 30-40º N were characterized by a second peak around 16-20 h LT [*Astafyeva et al.,* 2015; *Nava et al.,* 2016]. The main mechanism behind the ionospheric perturbation during this time intervals is, most probably, the prompt penetration of electric field (PPEF) [*Fagundes et al.,* 2016; *Nava et al.,* 2016; *Paul et al.*, 2018].

The TEC variations observed for the Iberian Peninsula area (see Figs. 2-3 and S7 in SI) are also show positive-negative ionospheric storm on March 17-18. The change of the sign of the TEC response to the geomagnetic storm is most clearly seen in variations of EOF1 (Fig. 2, see also Fig. S5 in SI): the amplitude of EOF1 on March 17 reaches the highest value for this month, and the amplitude of EOF1 on March 18 is even smaller than the EOF1 monthly mean level. We also observe the second daily peak on March 17, which is clearly seen in the variations of the EOF2 amplitude (Fig. 3, see also Fig. S6 (bottom) in SI).

The geomagnetic storm on June 22 (GS2, see Tab. S2 in SI) is thoroughly described in, e.g., *Astafyeva et al.* [2018], *Astafyeva et al.* [2016], *Astafyeva et al.* [2017], *Balasis et al.* [2018], *Liu et al.* [2015], *Ngwira et al.* [2018], *Paul et al.* [2018], *Pazos et al.* [2019], *Piersanti et al.* [2017], and *Singh and Sripathi* [2017]. This storm, similarly to the one in March, was related to the arrival of two CMEs associated with two solar flares (M-class) on June 18 and 21. On June 23 the recovery phase began but it was partly disrupted by a small CME arrived on June 25 which resulted in another small decrease of *Dst* [*Pazos et al.,* 2019] on June 26.

The ionospheric response to this storm is also consisted of the positive (June 22) and negative (June 23-24) phases observed in all latitudinal sectors. In the middle latitudes of the European-African sector TEC increased (relative to the undisturbed level) on June 22 during the evening, pre-sunset hours end even after the sunset [*Astafyeva et al.,* 2017]. On June 23 TEC started to decrease on both dayside and nightside of the Earth [*Astafyeva et al.,* 2017], and dual peaks in TEC diurnal variation were observed at the middle and low latitudes of the Northern Hemisphere [*Paul et al.*, 2018]. The main proposed mechanism for the ionospheric response to the geomagnetic storm was, again, PPEF [*Astafyeva et al.,* 2016; *Ngwira et al.,* 2018; *Paul et al.*, 2018] and the disturbance dynamo electric field (DDEF) on June 23 [*Paul et al.*, 2018].

Similar to the March event, the amplitude of the TEC daily cycle over IP significantly increases on June 22 and, again, on June 25, during the second *Dst* decrease (see Figs. 2-3 and S8 in SI). On the following days (June 23 and 26, respectfully) the amplitude of the daily variation decreased. These changes are reflected in the variations of the amplitude of Mode 1 (EOF1). Similar behavior but with smaller amplitude difference was also observed during the weak disturbance on June 8-11. Mode 2 of the IP TEC variations shows more complex behavior. Similar to the March event, its amplitude is high and EOF2 is positive on the 1st day of the main storm (June 22) and negative on the 2nd day of the main storm (June 23). On contrary, for the second *Dst* decrease on June 25, as well as for the weak disturbance on June 8-11, the amplitude of Mode 2 is high but EOF2 is negative changing to almost zero but positive on the following day.

On October 7-8 a storm with *Dst* = –124 nT and *Kp* = 6 was registered (GS4, Tab. S2 in SI). This storm was not associated with any CME and was probably caused by a coronal hole high-speed streams [*Matsui et al.,* 2016; *Pazos et al.,* 2019]. No significant ionospheric response to this storm was previously reported.

The variations of TEC observed over IP (see Figs. 2-3 and S9) during this storm can be identified as a negative storm: the amplitude of the daily TEC cycle on the 1st day of the storm was lower than during previous and following days and they are most clearly seen in variations of Mode 1 (EOF1). Another feature of TEC variation during this storm is that Mode 2 increased on both days of the storm causing a second peak in TEC daily cycle on October 7 and an offset of the daily maximum on October 8.

The last storm of 2015 started in the evening of December 19 and lasted until December 22 (GS3, see Tab. S2 in SI). *Dst* = –155 nT with *Kp* = 6 were observed on the night from December 20 to December 21. This was a two-step storm [*Balasis et al.,* 2018] caused by CME arrived on December 19 and related to a flare on December 16 [*Balasis et al.,* 2018] resulting in an auroral activity all over the northern Europe [*Cherniak and Zakharenkova,* 2018]. Detailed description of this storm can be found, e.g., in *Balasis et al.* [2018], *Cherniak and Zakharenkova* [2018], *Loucks et al.* [2017], and *Paul et al.* [2018].

During this geomagnetic storm the positive ionospheric storm was observed in the middle latitudes of the Asian sector on December 20-21 [*Paul et al.,* 2018] without the consequent negative phase.

This storm was characterized by the increase of the amplitude of the TEC daily variation over IP (see Figs. 2-3 and S10), as well as Mode 1, on the 1st day (positive storm). On the 2nd day the amplitude of the daily variations was still high but lower than on the 1st day. Probably this ionospheric storm can be classified as a positive-negative storm as well, and the relatively high amplitude of the TEC variations on the 2nd day is related to the two C-class flares that were observed on December 21. Mode 2 variations were significantly amplified only on the 1st day of the storm. The small geomagnetic disturbance on December 14-15 (GD3, DOY348-349) seems to have no effect on the TEC variations.

Among four strongest geomagnetic storms of 2015 (GS1-GS4) only one (GS3) was characterized by just the negative ionospheric storm. Two geomagnetic storms (GS1 and GS2) caused ionospheric disturbances that can be classified as positive-negative storms. Lastly, the geomagnetic storm GS4 resulted in the ionospheric storm that either positive or positive-negative. The types of the ionospheric storm defined by the original TEC values and by their Mode 1 are the same. Thus, the analysis of EOF1 allows easy classification of an ionospheric storm.

Besides the overall increase or decrease of the amplitude of the TEC variations described by Mode 1, there is another prominent feature of the TEC variations extracted from the data as Mode 2. Mode 2 is associated with appearance, as a rule, of a second daily peak or a sharp decrease in the TEC variations during the afternoon hours. Mode 2 was shown to intensify on the 1st day of the storm (causing second daily peak in TEC). This behavior is seen for all four storms of 2015. On the 2nd day of the storms in March and June the Mode 2 was also high in amplitude but of the opposite sign causing sharp decrease of the TEC in the afternoon hours. During the storm in October this mode was both significant in amplitude and of the same sign as for the previous days resulting, however, not in second peak but in the offset of the daily peak to the afternoon hours. During the December storm the amplitude of Mode 2 on the 2nd day was negligible.

Among the geomagnetic disturbances with lower amplitudes (GD1-GD3), GD1 and GD2 caused positive-negative ionospheric disturbances. The amplitude of the Mode 2 was high on the 1st day of the disturbances with positive EOF2 for GD2 and negative one for GD1. On the 2nd day of both disturbances EOF2s changed their signs to opposite. The GD3 in December showed no effect on the ionospheric TEC variations.

The mean correlation coefficients between the EOF1s and EOF2s and the geomagnetic indices (*Dst*, *Kp* and $K_{COI}$, *ap* and *AE*) for each analyzed month are shown in Tab. 1. The correlation coefficients were calculated with lag = 0-2 days (geomagnetic indices lead). EOF1s anti-correlate with variations of the *K* and *AE* indices with lag = 1 day for all months except December (lag = 0 days). EOF2s anti-correlate with variations of the *K* and *AE* indices for March and June (lag = 1-2 days) and correlate for October and December (lag = 0 days). For the *Dst* index the signs of the correlation coefficients are opposite. The highest (in the absolute value) correlation coefficients were obtained, as a rule, for the *Dst* and *AE* indices. The change of the sign of the correlation coefficients between different events reflects, to our mind, differences in the geomagnetic conditions of the studied months, non-linearity of the relation between the amplitude of the Modes 1 and 2 and geomagnetic activity, and differences in the ionospheric response to different storms described above.

Previously, some conclusions were made about dependence of ionospheric response to geomagnetic storms on the season [*Andonov et al.,* 2011; *Kutiev et al.*, 2013 and references therein] or starting time [*Borries et al.*, 2015; *Yamamoto et al.*, 2000; see also *Kutiev et al.,* 2013 and references therein]. However, since only seven geomagnetic disturbances appeared during different seasons and started at different time were considered in this work, we can make no definite conclusion about dependence of TEC variations during geomagnetic storms on the season or starting time.

We also calculated correlation coefficients between TEC EOFs and the parameters of the solar wind (Tab. 1) with different lags (solar wind parameters lead). Both EOF1 and EOF2 show high and statistically significant correlations with solar wind parameters with highest (in the absolute value) correlation coefficients for *B, Bx*, *Bz*, *v* and *p*.

5.2 Solar flares, UV and XR fluxes

The X-flare on March 11 had very weak effect on the TEC variations over IP: there is weak increase of TEC at 17 h (first hourly measurements after the flare). Neither EOF 1 nor EOF 2 show particular response to this flare (see Figs. 2-3).

On the time scale of a month there is dependence between the number of solar flares, solar UV and XR fluxes, and the amplitude of the daily TEC cycle and the amplitude of the Mode 1 (or EOF1). The correlation coefficients (see Tab. 1 and its extended version Tab. S3 in SI) between the EOF1s, and the solar UV and XR proxies are r = 0.2-0.86 (depending on the month and the proxy). Generally, the UV flux proxies have higher correlation coefficients than the XR proxy: in average $r = 0.57$ for EOF1 vs UV and $r = 0.46$ for EOF1 vs XR proxies. The lowest correlation is found for December. The correlation coefficients between the EOF2s, and the solar UV and XR proxies are low and statistically insignificant. Thus, the overall increase of the flares number and/or solar UV and XR fluxes results in the increase of the amplitude of the daily TEC variation and, consequently, EOF 1. The dependence of EOFs on the number of flares is weaker and varies with month.

### 6 MRM of TEC response to space weather forcing

The results of the correlation analysis presented in Section 5 show relatively high correlations between the TEC series and space weather parameters: the number of the solar flares, UV and XR fluxes, solar wind parameters and geomagnetic indices. Consequently, linear multiple regression method can be used to reconstruct TEC variations using space weather parameters as regressors. Since the purpose of these multiple regression models (MRM) is not to prove physical relations between the ionospheric and space weather parameters but to obtain best prediction quality possible, we can use as regressors parameters that could be related between each other (both through physical processes and statistically). Similar, but not exactly the same, approaches were taken in *Chen et al.* [2015] and *Li et al.* [2019] to model TEC variations over large areas in dependence on the solar (F10.7 index) and geomagnetic (Ap index) activity. Also, in the recently published paper by *Razin and Voosoghi* [2020] the first three PCA modes were used to train artificial neural networks to predict TEC variations.

The MRMs were constructed for EOF1 and EOF2 series as well as for the TEC daily means (1 d TEC series). Then, the MRM fits for the EOF1, EOF2 and 1 d TEC were used, together with previously obtained PC1 and PC2, to reconstruct TEC series (PCA-MRM reconstruction). The advantage of this method is that we do not need to parameterize daily TEC variations and their dependence on the season since these daily variations are already extracted from the data as corresponding PC1 and PC2. The MRM were constructed using time lags from 0 (no lag) to 2 days (space weather parameters lead by 2 days). Using the lagged MRM we test the possibility to use this method to forecast TEC variations.

We constructed MRMs both for the mean TEC series and for the series from the individual receiver (*SCI-LIS*). Since we make MRMs for the series with 1 d time resolution, we used daily mean series of space weather parameters too. Two types of MRM were constructed. The models of the first type use as regressors the solar (i.e., fluxes and flares) and geomagnetic parameters and the second type models use as regressors the solar and solar wind parameters. These MRMs were constructed using the "best subset" method to study which parameters are most/less frequently used for best fitting models (models with highest possible for the current set of regressors value of $R_{adj}^2$). The $R_{adj}^2$ parameters for MRMs with the solar wind parameters as regressors are shown in Tab. 2 (and for both sets of regression in Tab. S4 in SI) for the mean TEC series and in Tab. 3 for the *SCI-LIS* TEC series. The correlation coefficients between the original mean TEC series and the PCA-MRM reconstructions are also shown in Tabs. 2-3 and Tab. S4 in SI. In most cases the prediction quality of the MRMs using solar wind parameters as regressors are higher than the ones using geomagnetic indices. Solar UV flux proxy (*Mg II*) as well as the interplanetary magnetic field components $Bx$ and $Bz$ are the parameters most often chosen ($> 43\%$ of the MRMs use them) for "best set" of the regression parameters for the mean TEC series. For MRMs of the *SCI-LIS* TEC series the most often selected regressors are solar UV flux proxy (*Mg II*) and the solar wind flow speed $v$ ($> 75\%$ of the MRMs use them), as well as the interplanetary magnetic field components $B$, $Bx$ and $By$ (58-67% of the MRMs use them). For most of the PCA-MRM reconstruction the correlation coefficients between the original and reconstructed series are higher for the individual receiver than for the mean TEC series, however, in many cases the MRM reconstructions of the daily mean TEC or EOFs for the mean TEC series are better than for the individual receiver (please compare Tabs. 2 and 3).

The PCA reconstructions using original EOFs for the *SCI-LIS* TEC series are shown in Fig. 4 and the PCA-MRM reconstructions using regression on the solar wind parameters with lag = 2 days are shown in Fig. 5. The corresponding results for lag = 0-1 days are in SI, Figs. S11-S12. The two first PCA modes allows to reconstruct 90-95% of the original TEC series variability (Fig. 4). In case a higher reconstruction quality is needed we suggest to add the 3rd mode. The PCA-MRM reconstructions of the original TEC series (Fig. 5) have lower quality: 52-92% of the observed variations are reconstructed (Tab. 2), depending on the month and the lag.

To test forecasting abilities of the PCA-MRM reconstruction method we used the data from the receiver *SCI-LIS*. First, we constructed MRM models for the daily mean TEC and EOFs using the data for a selected number of the days of a month (hereafter, "training days"). The minimal length of the "training days" is 16 for all months except December (due to statistical properties of the regressors' data sets the minimal length of the "training days" for December is 20 days). We used these MRM reconstructions to forecast the TEC variations for the following day, and then we gradually increased the length of the "training days". To make reconstruction technically easier we did not applied the "best subset" selection; therefore, the constructed MRMs may not be the ones with the highest $R_{adj}^2$ values. We made forecasts using the space weather parameters lagged by 1 and 2 days and also calculated their mean. Some examples of the forecasts based on different length of the "training days" for March 2015 are shown in Fig. 6 and for other months are shown in Figs. S13-S15 (SI). To our mind the use of the "best subset" and combination of different lags for different predictors (e.g., smaller lags for the solar flux proxies and larger for the solar wind parameters) could improve the forecast quality.

The quality of the PCA-MRM forecasting was estimated by calculation of the following statistics: correlation coefficients $r$, $RMSE/\sigma$ and $\chi^2$ between the observed and forecasted series (see Fig. 7 for March and Figs. S16-S18 for June, October and December, respectively). In general, the quality of the TEC forecast growths with the number of the "training days". The forecasting quality ($r$, $RMSE/\sigma$ and $\chi^2$) decreases for days of geomagnetic storms storm: e.g., $r$ ($RMSE/\sigma$) is significantly low (high) on March 17-18 (Fig. 7, see also Fig. 6. top left) and June 22 (Fig. S16, see also Fig. S13, bottom left in SI) – days of the most severe storms of 2015. The MRMs based on a set of days that already includes a storm have better forecasting abilities. Also, $\chi^2$ gradually decreases to 1 when the number of the "training days" is increased. Averaging of the forecasts with different lags, in general, improve the forecast quality. We plan to test other regression methods, including neural networks based on the machine learning algorithms, especially for the storm days, to find an optimal set of the regressors and number of training days, to test different time lags and their combination, to study seasonal variations of the forecasting quality and make other optimizations. However, to our mind, the results of this exploratory work show good potential for the PCA-MRM forecast method. We also found that for the region as IP the spatial TEC variations are not essential for forecasting daily TEC variations; therefore both the TEC series from the individual receiver and mean TEC series can be used.

## 7 Conclusions

In this work we analyzed variations of the total electron content (TEC) over the mid-latitudinal area of the Iberian Peninsula (IP) during seven geomagnetic disturbances of 2015 with $Dst < -50$ nT that occurred in March, June, October and December. We also analyzed variations of TEC in relation to the daily total number of solar flares (C, M and X classes) and the solar UV and XR fluxes.

The TEC data were obtained for two locations on the Iberian Peninsula: Lisbon (Portugal) on the west coast and Ebro (Spain) on the north-east coast of the peninsula. Two types of TEC were used in the study: TEC calculated from GNSS receivers and the ionospheric TEC (iTEC) obtained from the ionosonde at the Ebro Observatory. All TEC series are in good agreement. The new data series, the non-calibrated TEC from the SCINDA GNSS receiver installed between 2014 and 2019 in the Lisbon airport area, is found to be well correlated with other series which allowed us to use it for the analysis.

The TEC series were submitted to the principal component analysis allowing to extract two main modes that together explain > 90% of the TEC variability. The Mode 1 represents daily TEC variation and the Mode 2 is associated with either second daily peak or a sharp decrease during the afternoon hours (19-21 h LT).

Similar changes in the amplitude of the TEC daily variation and EOF1 were observed during all studied geomagnetic disturbances. Thus, Mode 1 (or corresponding EOF1) can be used for easy classification of an ionospheric storm.

Mode 2 of the TEC variations during most of the studied storms increased in amplitude with positive EOF2 values on the $1^{st}$ day of the storm and negative (or almost zero in case of December storm) EOF2 values on the $2^{nd}$ day. Thus, the high positive Mode 2 (or high positive EOF2) is a good indicator of a second daily peak in the TEC daily variations.

The overall increase/decrease of the flares number as well as changes of the solar UV and XR fluxes during the analyzed months resulted in the increase/decrease of the amplitude of the TEC daily cycle (and therefore Mode 1). No relation between the amplitude of the Mode 2 associated with the solar UV and XR fluxes was found.

The amplitude of the Mode 1 and Mode 2 (EOF1 and EOF2) were found to correlate with space weather parameters: solar UV and XR fluxes, daily numbers of the flares, parameters of the solar wind and geomagnetic indices. Based on the correlation analysis we constructed multiple regression models (MRM) for the EOFs and the daily mean TEC values with space weather parameters as regressors. MRMs were made with time lags = 0-2 days (space weather parameters lead). The MRM fits for EOFs and the daily mean TEC values were used to reconstruct TEC variations using corresponding PCs (PCA-MRM reconstructions).

We also explored a possibility to use the PCA-MRM reconstruction method to make forecasts of the daily TEC variations. In general, the predictive quality of the PCA-MRM forecasts were found good but further studies are needed to improve their predictive quality.

## Acknowledgments, Samples, and Data

We thank anonymous Reviewer for helpful comments.

CITEUC is funded by National Funds through FCT - Foundation for Science and Technology (project: UID/MULTI/00611/2019) and FEDER – European Regional Development Fund through COMPETE 2020 – Operational Programme Competitiveness and Internationalization (project: POCI-01-0145-FEDER-006922).

This research was supported through the project "SWAIR - Space weather impact on GNSS service for Air Navigation", ESA Small ARTES Apps, https://goo.gl/YN2iJf.

We are grateful to SEGAL (Space and Earth Geodetic Analysis laboratory) and personally to Dr. Rui Fernandes from University of Beira Interior (Portugal) for the access to the SCINDA receiver data. The *SCI-LIS* TEC data are available at Barlyaeva, T., Barata, T., Morozova, A., 2020. Datasets of ionospheric parameters provided by SCINDA GNSS receiver from Lisbon airport area, Mendeley Data, v1 http://dx.doi.org/10.17632/kkytn5d8yc.1

"SCINDA-Iono" toolbox for MATLAB by T. Barlyaeva is available online at https://www.mathworks.com/matlabcentral/fileexchange/71784-scinda-iono_toolbox.

We acknowledge the mission scientists and principal investigators who provided the data used in this research:

The TEC *ROB-LIS* and *ROB-EBR* data sets are from the Royal Observatory of Belgium (ROB) data base and are publicly available in IONEX format at ftp://gnss.oma.be/gnss/products/IONEX/, see also *Bergeot et al. [2014]* for more information.

We also wish to thank the Ebro Observatory (OE, Univ. Ramon Llull- CSIC) and Dr. Germán Solé for providing ionosonde data (TEC *IONO-EBR*), http://www.obsebre.es/en/ionospheric-data-catalogs .

We acknowledge the use of the *Dst* index from the Kyoto World Data Center http://wdc.kugi.kyoto-u.ac.jp/dst_final/index.html.

Geomagnetic data measured by the OGAUC are available at the World Data Centre for Geomagnetism web portal http://www.wdc.bgs.ac.uk/dataportal/ .

We acknowledge the use of the *Kp* index from the GFZ German Research Centre for Geosciences https://www.gfz-potsdam.de/en/kp-index/.

The solar wind data and the ap index are from the SPDF OMNIWeb database. The OMNI data were obtained from the GSFC/SPDF OMNIWeb interface at https://omniweb.gsfc.nasa.gov, see also *King and Papitashvili [2004]* for more details.

The F10.7 index was also obtained from the OMNI data base at https://omniweb.gsfc.nasa.gov/form/dx1.html.

The Mg II data are from Institute of Environmental Physics, University of Bremen http://www.iup.uni-bremen.de/gome/gomemgii.html, see also *Snow et al. [2014]* for more information.

The data on the variations of the solar XR flux are from the LASP Interactive Solar Irradiance Data Center (LISIRD, http://lasp.colorado.edu/lisird/ ). LISIRD provides a uniform access interface to a comprehensive set of Solar Spectral Irradiance (SSI) measurements and models from the soft X-ray (XUV) up to the near infrared (NIR), as well as Total Solar Irradiance (TSI). The XR$_{TIMED}$ data are from the Solar EUV Experiment (SEE) measures the solar ultraviolet full-disk irradiance for the NASA TIMED mission. Level 3 data represent daily averages and are filtered to remove flares available at http://lasp.colorado.edu/lisird/data/timed_see_ssi_l3/.

The X-ray Flare dataset was prepared by and made available through the NOAA National Geophysical Data Center (NGDC). The data about the solar flares for 2015 are from https://www.ngdc.noaa.gov/stp/space-weather/solar-data/solar-features/solar-flares/x-rays/goes/xrs/goes-xrs-report_2015_modifiedreplacedmissingrows.txt.

**Figure captions**

**Figure 1.** Mean TEC PC1 (top) and PC2 for March (green), June (red), October (orange) and December (blue).

**Figure 2.** Mean TEC EOF1 for March (a), June (b), October (c) and December (d). Shaded areas mark geomagnetic storms.

**Figure 3.** Same as Figure 2 but for EOF2.

**Figure 4.** *SCI-LIS* TEC series (black lines) and their PCA reconstructions (red lines) for March (a), June (b), October (c) and December (d).

**Figure 5.** Same as Figure 4 but for PCA-MRM reconstructions with space weather parameters (solar UV, flares and solar wind parameters) leading by 2 days for March (a), June (b), October (c) and December (d).

**Figure 6.** *SCI-LIS* TEC March series (black lines) and their PCA-MRM forecasts using space weather data with lag = 1 day (red lines), lag = 2 days (blue lines) and their mean (green lines). Different number of the "training days" is used (16, 21, 25 and 30 days). The forecasted day is shaded grey.

**Figure 7.** Quality of the PCA-MRM forecast in dependence on the number of the "training days" for March 2015: (a) correlation coefficients between the observed and forecasted TEC series, (b) RMSE divided by the standard deviation, (c) $\chi^2$ calculated for the 1 h (solid) and 1 d series (dashed) for MRM with lag = 1 (red), lag = 2 (blue) and their mean (green).

**Table 1.** Correlation coefficients *r* between EOFs and space weather parameters. Only $|r| \geq 0.2$ and *p value* $\leq 0.2$ (in parentheses) are shown. The lag in days is shown in brackets (space weather parameters lead). Bold marks highest (in absolute values) correlation coefficients for each month for each group of parameters. The full version of Table 1 is in SI (Table S3).

|        | March | June | October | December |
|--------|-------|------|---------|----------|
| EOF1   |       |      |         |          |
| Mg II  | **0.46 [1/2] (0.03)** | **0.74 [0] (0.01)** | **0.91 [1] (0.03)** | 0.26 |
| Dst    | **0.56 [1] (<0.01)** | **0.51 [0/1] (0.05)** | **0.75 [0/1] (0.14)** | -0.63 [0] (<0.01) |
| Kp     | -0.43 [1/2] (0.05) | -0.35 [1/2] (0.18) | -0.64 [1/2] (0.17) | **0.76 [0] (<0.01)** |
| AE     | **-0.55 [1] (<0.01)** | -0.35 [1] (0.16) | **-0.74 [1] (0.09)** | **0.77 [0] (<0.01)** |
| ap     | -0.43 [2] (0.04) | -0.45 [1] (0.03) | -0.57 [1/2] (0.014) | **0.81 [0] (<0.01)** |
| B      | **0.55 [0] (<0.01)** | **-0.43 [1] (0.03)** | -0.34 [2] (0.12) | 0.67 [0] (<0.01) |
| Bx     | -0.44 [0] (0.05) | 0.33 [2] | **0.59 [1/2] (0.01)** | -0.21 [2] |
| By     |  |  | -0.54 [0/1] |  |
| Bz     | **0.57 [1] (<0.01)** | 0.24 [1] | 0.46 [0/1] | **-0.70 [0] (<0.01)** |
| n      | 0.46 [0] (0.06) | **0.51 [0] (<0.01)** | 0.51 [0] (<0.01) | 0.54 [0] (<0.01) |
| v      | -0.40 [1] (0.1) | -0.23 [1] | **-0.60 [1] (0.14)** |  |
| p      | **0.53 [0] (<0.01)** | **0.51 [0] (<0.01)** |  | **0.60 [0] (<0.01)** |
| EOF2   |       |      |         |          |
| Mg II  |  | 0.22 | -0.25 |  |
| Dst    | 0.34 [1] (0.10) | **0.42 [2] (<0.01)** | -0.62 [0] (0.07) | -0.52 [0] (<0.01) |
| Kp     | -0.35 [1] (0.09) | **-0.4 [2] (<0.01)** | **0.69 [0] (<0.01)** | 0.61 [0] (<0.01) |
| AE     | **-0.44 [1] (0.02)** | -0.37 [0/2] (0.06) | **0.71 [0] (<0.01)** | **0.72 [0] (<0.01)** |
| ap     | **-0.41 [1] (0.02)** | **-0.46 [2] (<0.01)** | 0.52 [0] (0.07) | **0.78 [0] (<0.01)** |
| B      | 0.50 [<0.01] () | -0.26 [2] (0.08) | 0.33 [0] (0.17) | 0.71 [0] (<0.01) |
| Bx     |  | **0.49 [1] (<0.01)** |  |  |
| By     |  | -0.24 [2] (0.17) | 0.37 [2] | -0.20 [2] |
| Bz     | **0.54 [1] (<0.01)** | 0.37 [2] (0.03) | **-0.51 [0/1] (0.15)** | **-0.76 [0] (<0.01)** |
| n      | 0.35 [0] (0.08) | 0.21 [0] (0.16) | 0.20 [0] | 0.84 [0] (<0.01) |
| v      | -0.25 [1] | **-0.30 [0] (0.02)** | 0.39 [0] | -0.50 [2] (0.01) |
| p      | **0.45 [0] (<0.01)** | -0.20 [1] | **0.42 [0] (<0.01)** | **0.77 [0] (<0.01)** |

**Table 2.** Parameters of MRM for the daily mean TEC, EOF1 and EOF2 for different months on the solar UV, solar flares and solar wind parameters: $R_{adj}^2$ – the part of the variability of the independent parameter explained by MRM with adjustment for the number of degrees of freedom; r(TEC$_{MRM}$ vs TEC$_{orig}$) – correlation coefficients between the original TEC series and the PCA reconstructions of TEC using corresponding MRMs; the lag between the space weather and TEC series is between 0 and 2 days (space weather parameters lead).

| | March | June | October | December |
|---|---|---|---|---|
| lag = 0 | | | | |
| daily mean TEC $R_{adj}^2$ | 0.78 | 0.75 | 0.88 | 0.73 |
| EOF1 $R_{adj}^2$ | 0.79 | 0.74 | 0.88 | 0.74 |
| EOF2 $R_{adj}^2$ | 0.42 | 0.15 | 0.58 | 0.83 |
| r(TEC$_{MRM}$ vs TEC$_{orig}$) | 0.89 | 0.89 | 0.62 | 0.88 |
| lag = 1 | | | | |
| daily mean TEC $R_{adj}^2$ | 0.68 | 0.78 | 0.88 | 0.37 |
| EOF1 $R_{adj}^2$ | 0.68 | 0.78 | 0.88 | 0.37 |
| EOF2 $R_{adj}^2$ | 0.34 | 0.32 | 0.48 | 0.40 |
| r(TEC$_{MRM}$ vs TEC$_{orig}$) | 0.86 | 0.90 | 0.72 | 0.71 |
| lag = 2 | | | | |
| daily mean TEC $R_{adj}^2$ | 0.68 | 0.65 | 0.83 | 0.17 |
| EOF1 $R_{adj}^2$ | 0.69 | 0.65 | 0.83 | 0.17 |
| EOF2 $R_{adj}^2$ | 0.24 | 0.43 | 0.47 | 0.22 |
| r(TEC$_{MRM}$ vs TEC$_{orig}$) | 0.88 | 0.84 | 0.72 | 0.54 |

**Table 3.** Same as Table 2 but for the *SCI-LIS* TEC series.

| | March | June | October | December |
|---|---|---|---|---|
| lag = 0 | | | | |
| daily mean TEC $R_{adj}^2$ | 0.76 | 0.75 | 0.90 | 0.78 |
| EOF1 $R_{adj}^2$ | 0.57 | 0.32 | 0.78 | 0.51 |
| EOF2 $R_{adj}^2$ | 0.40 | 0.16 | 0.62 | 0.78 |
| r(TEC$_{MRM}$ vs TEC$_{orig}$) | 0.89 | 0.89 | 0.96 | 0.91 |
| lag = 1 | | | | |
| daily mean TEC $R_{adj}^2$ | 0.68 | 0.79 | 0.90 | 0.45 |
| EOF1 $R_{adj}^2$ | 0.46 | 0.29 | 0.70 | 0.44 |
| EOF2 $R_{adj}^2$ | 0.39 | 0.43 | 0.49 | 0.60 |
| r(TEC$_{MRM}$ vs TEC$_{orig}$) | 0.86 | 0.91 | 0.96 | 0.76 |
| lag = 2 | | | | |
| daily mean TEC $R_{adj}^2$ | 0.71 | 0.61 | 0.84 | 0.39 |
| EOF1 $R_{adj}^2$ | 0.41 | 0.35 | 0.56 | 0.44 |
| EOF2 $R_{adj}^2$ | 0.28 | 0.40 | 0.52 | 0.33 |
| r(TEC$_{MRM}$ vs TEC$_{orig}$) | 0.90 | 0.83 | 0.93 | 0.72 |

**Figure7.**

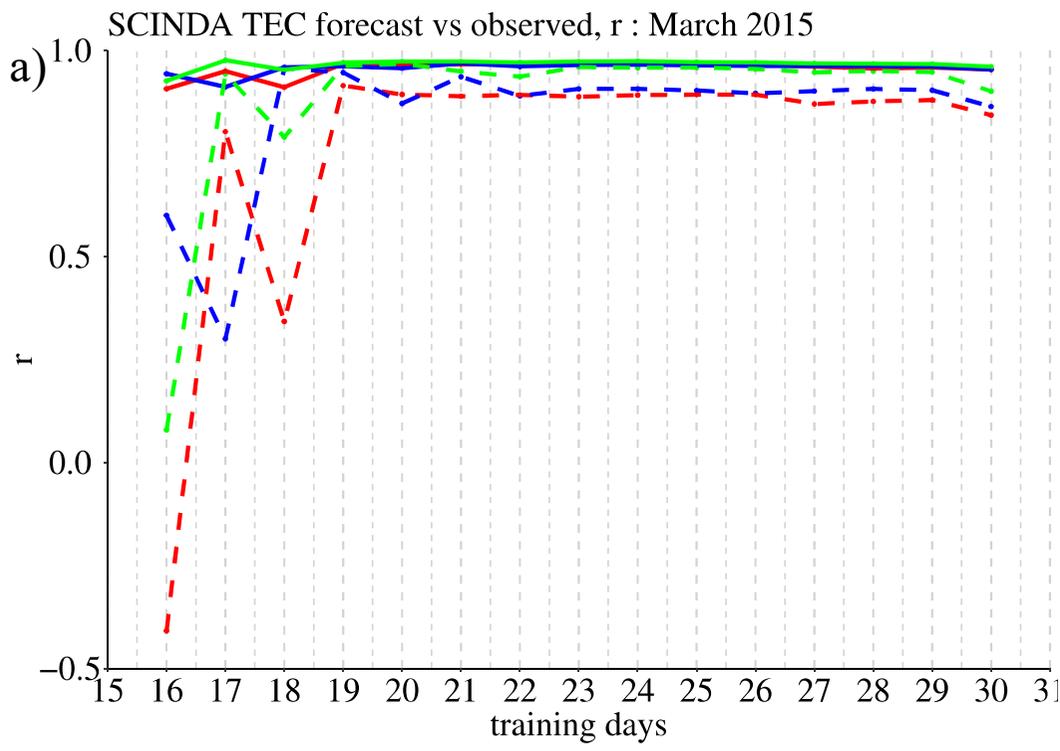

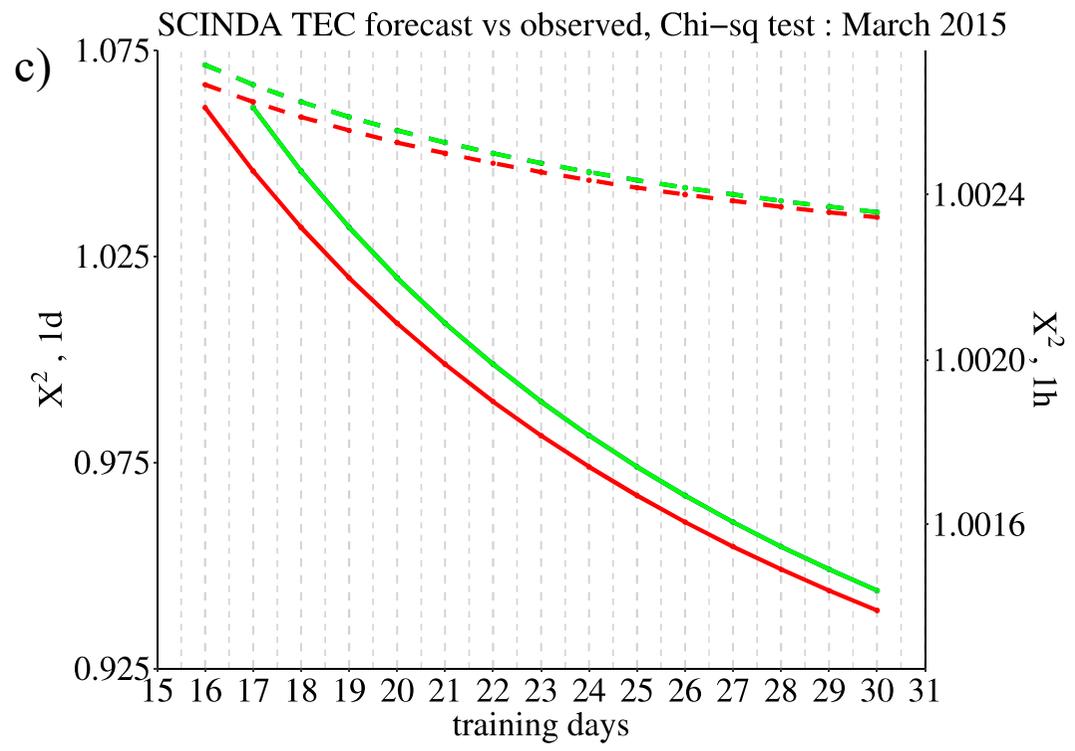

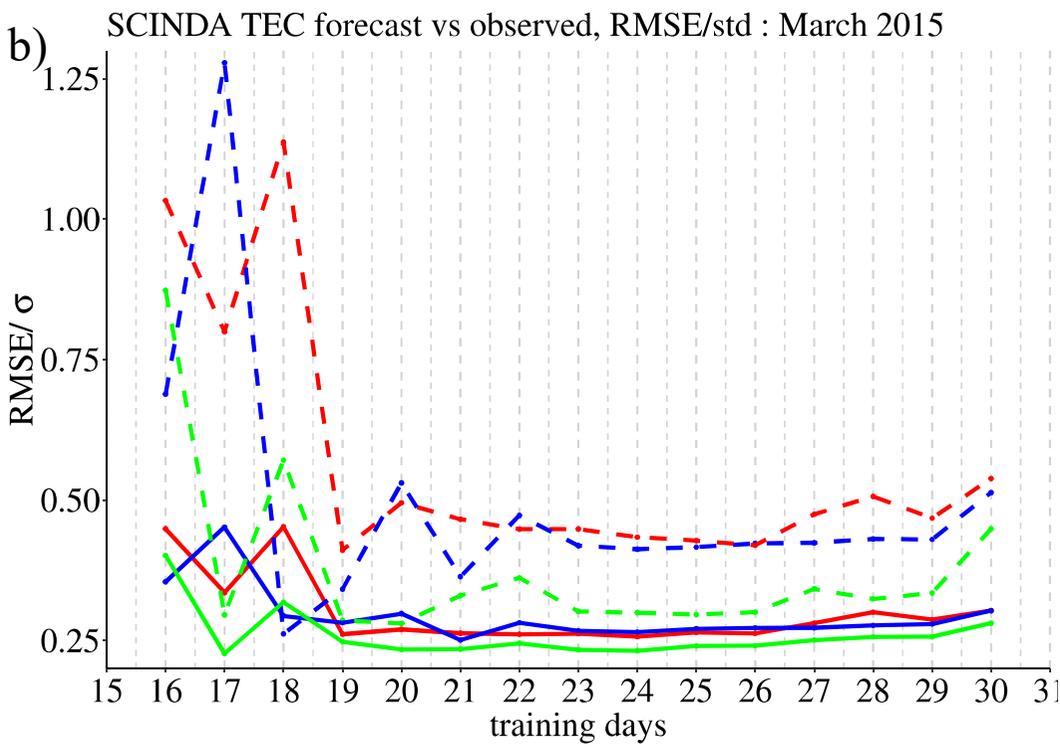

**Figure6.**

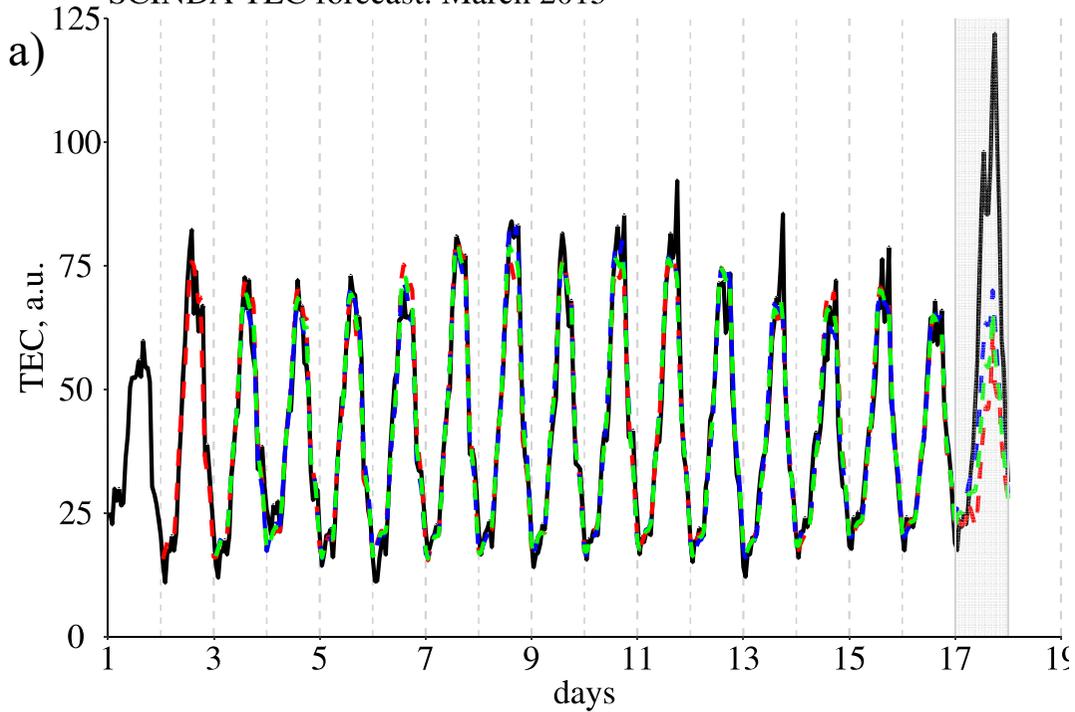

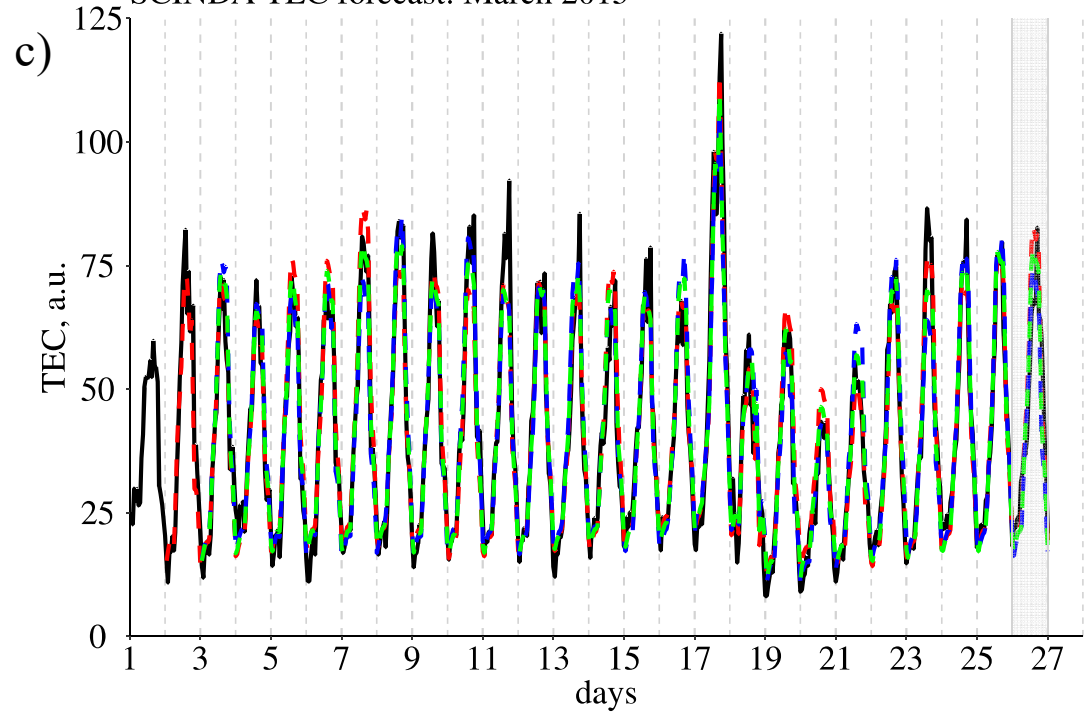

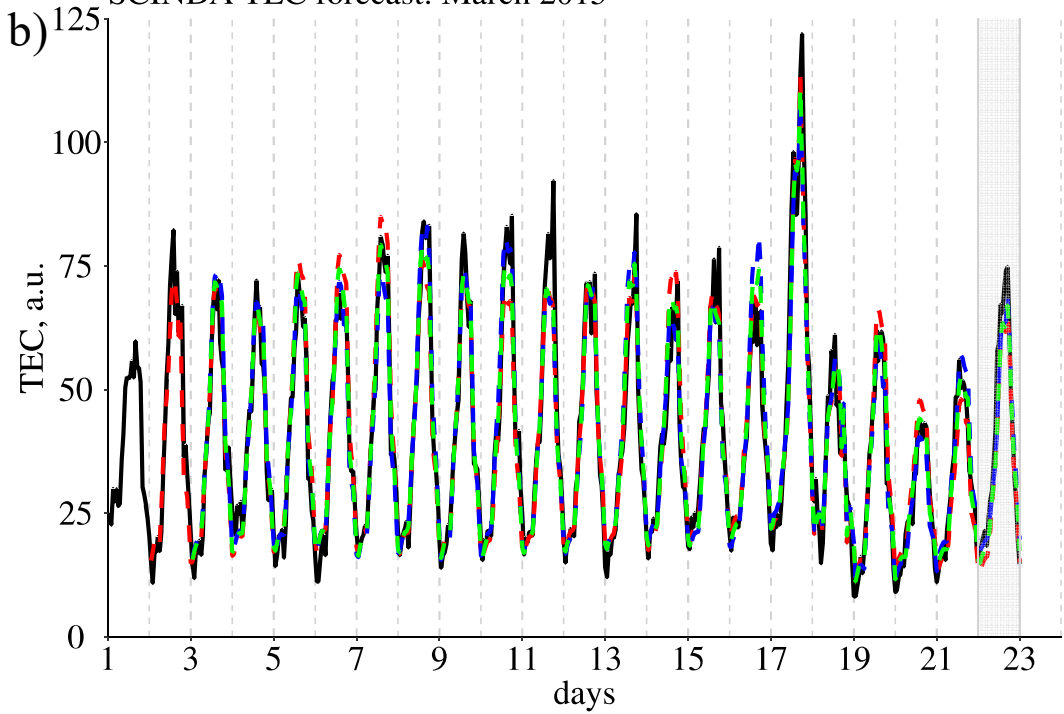

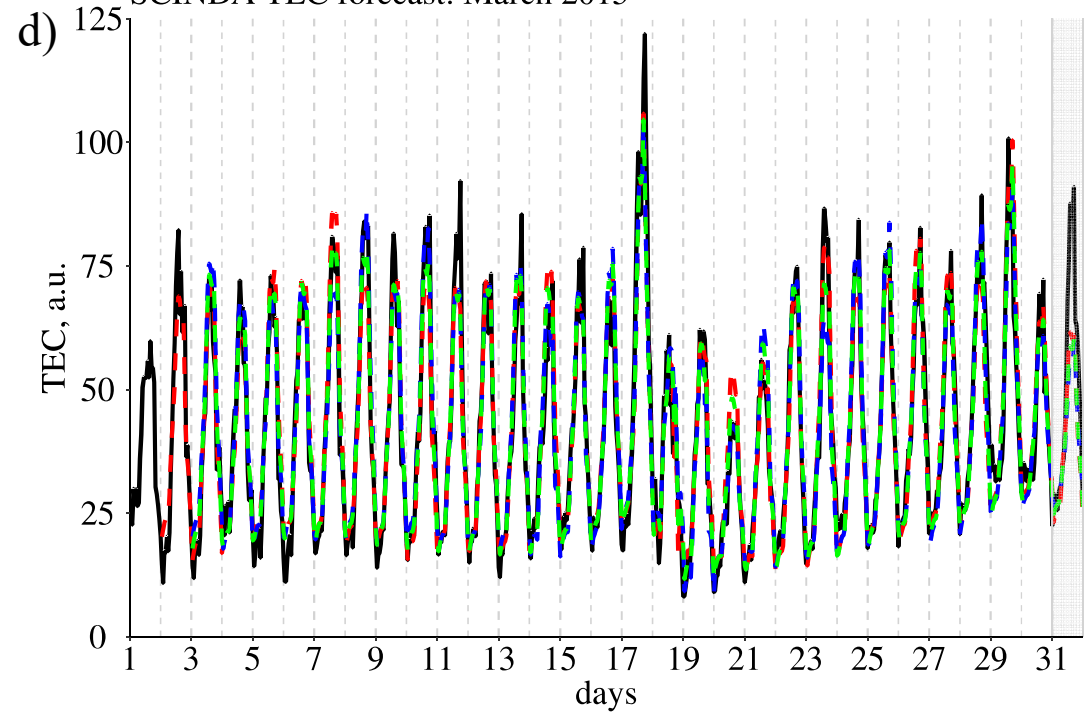

**Figure5.**

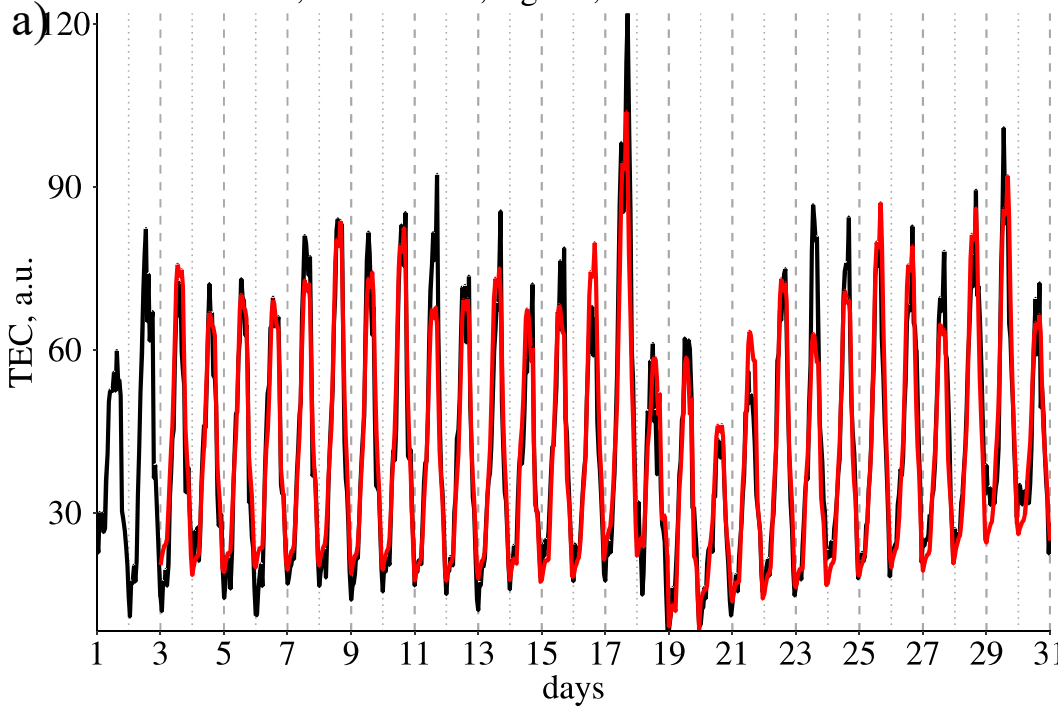

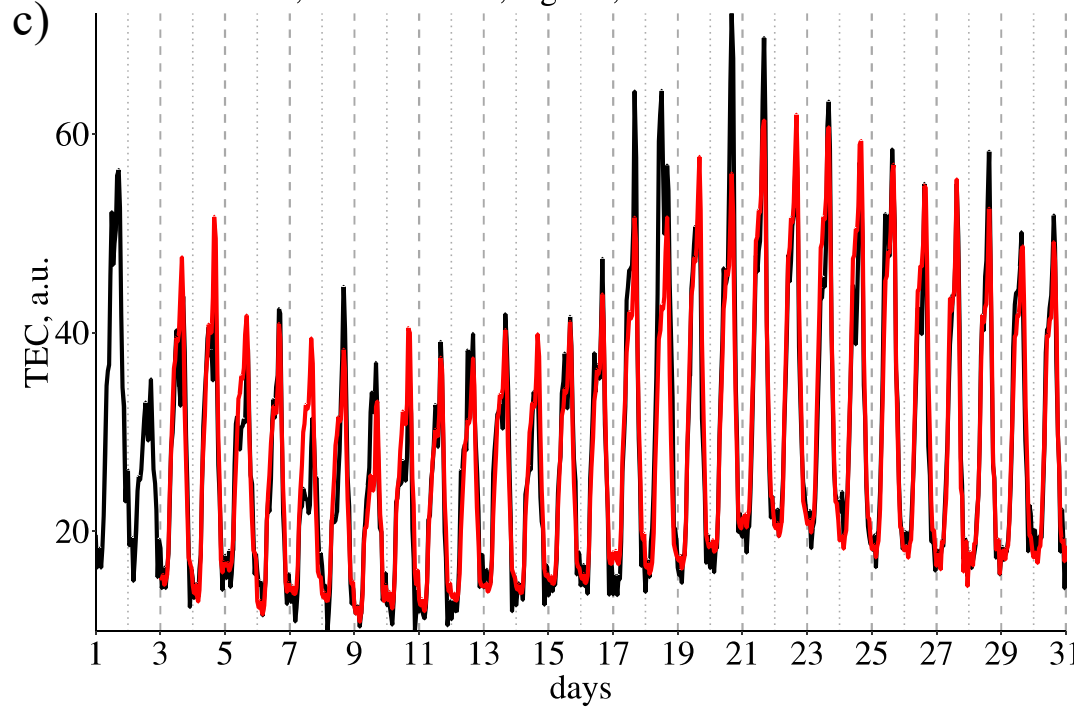

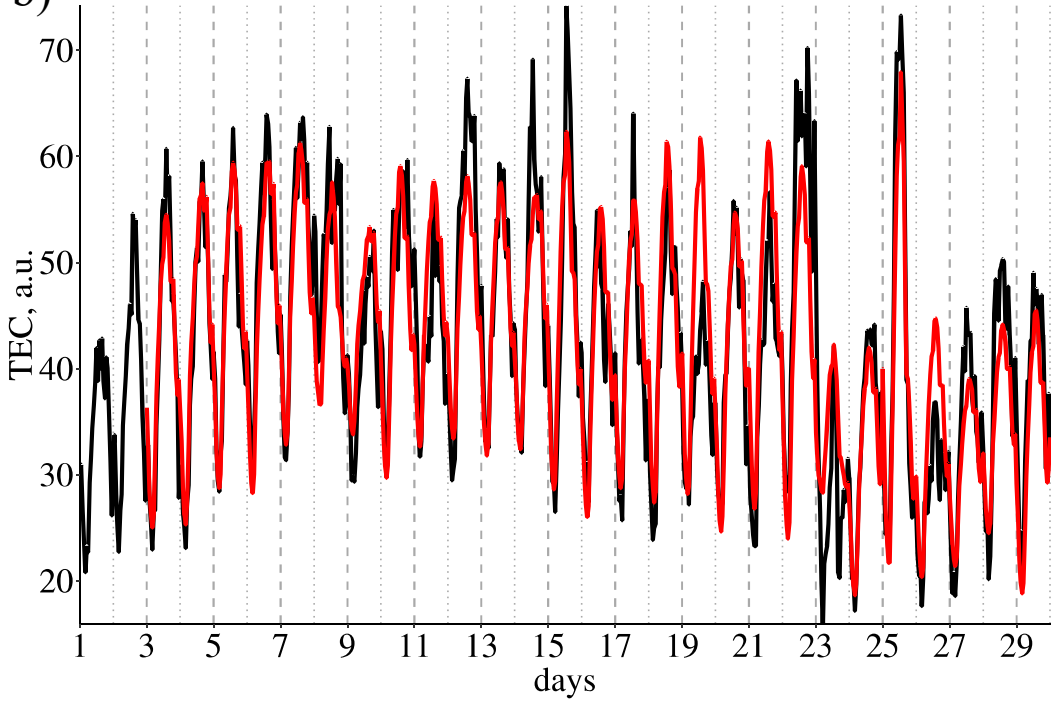

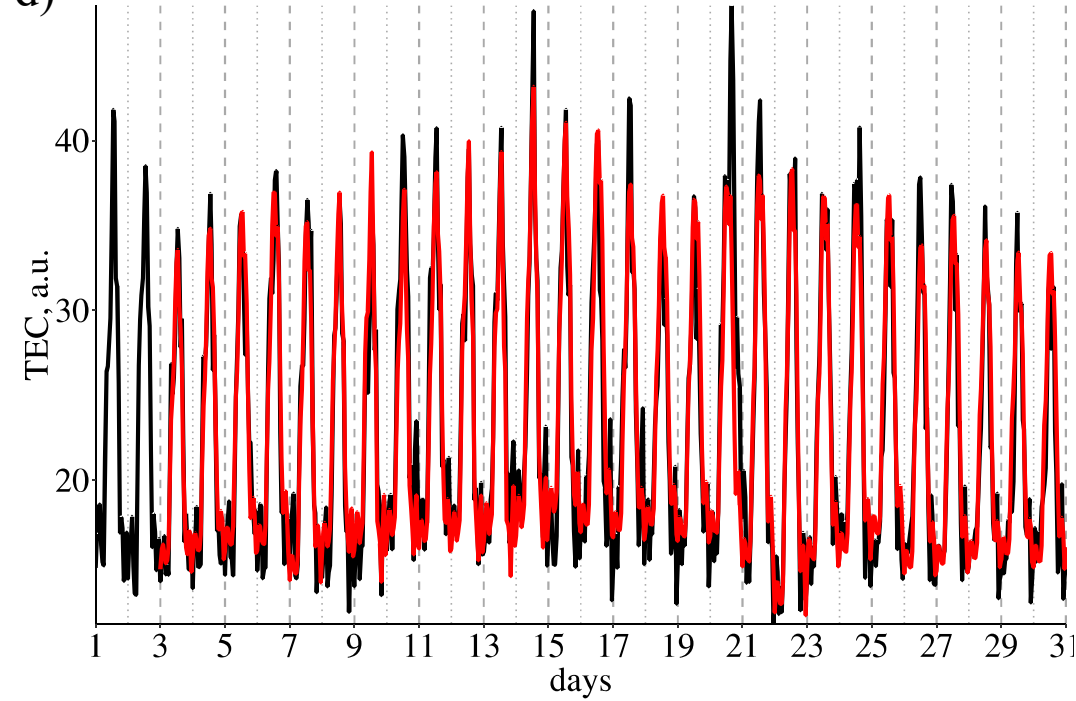

**Figure4.**

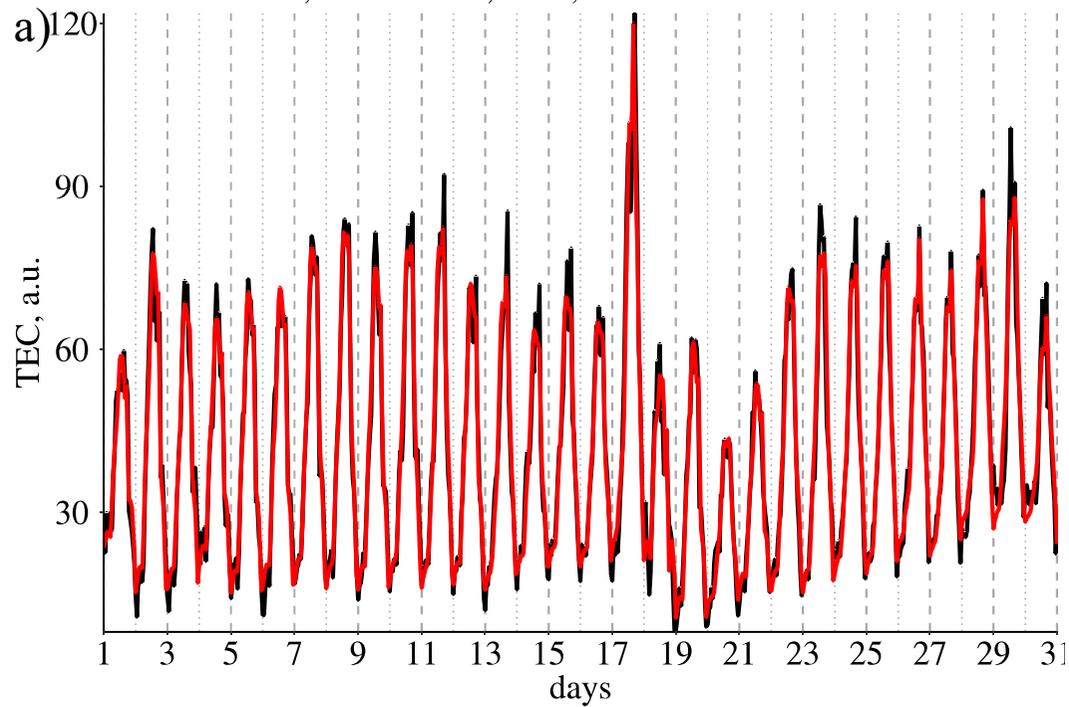

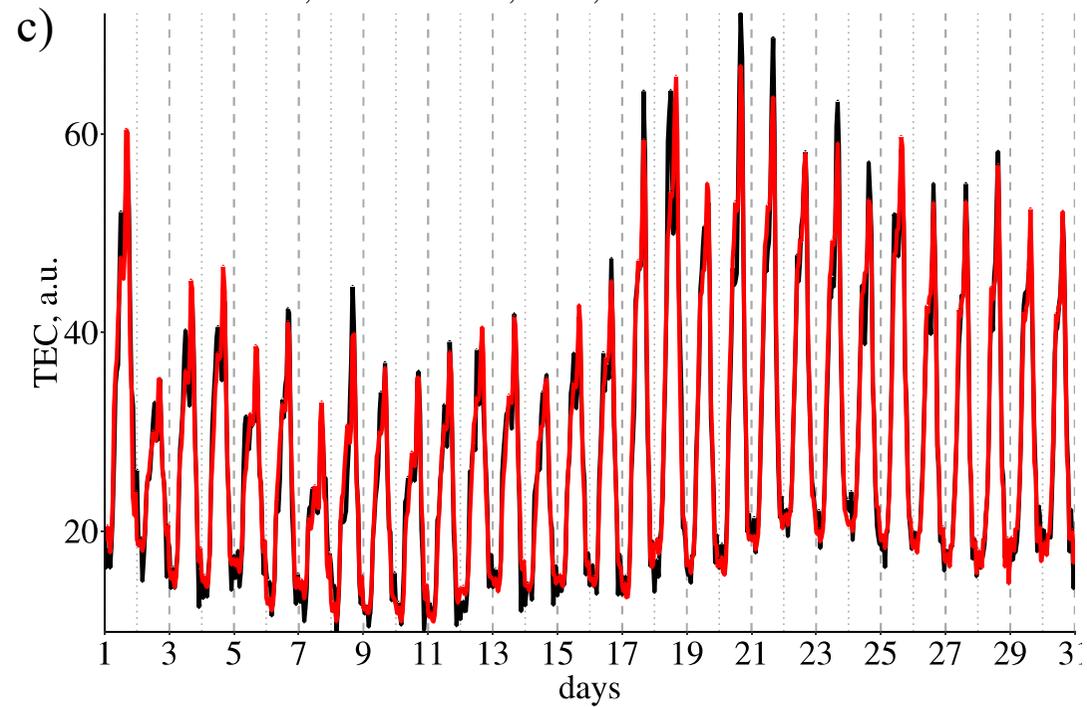

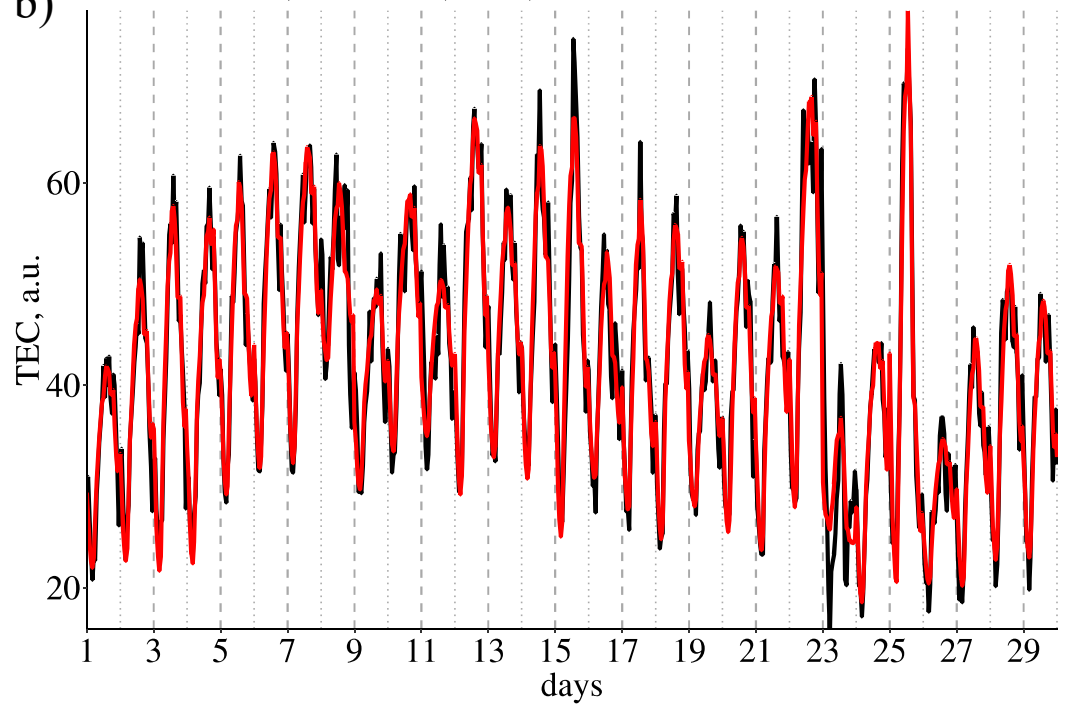

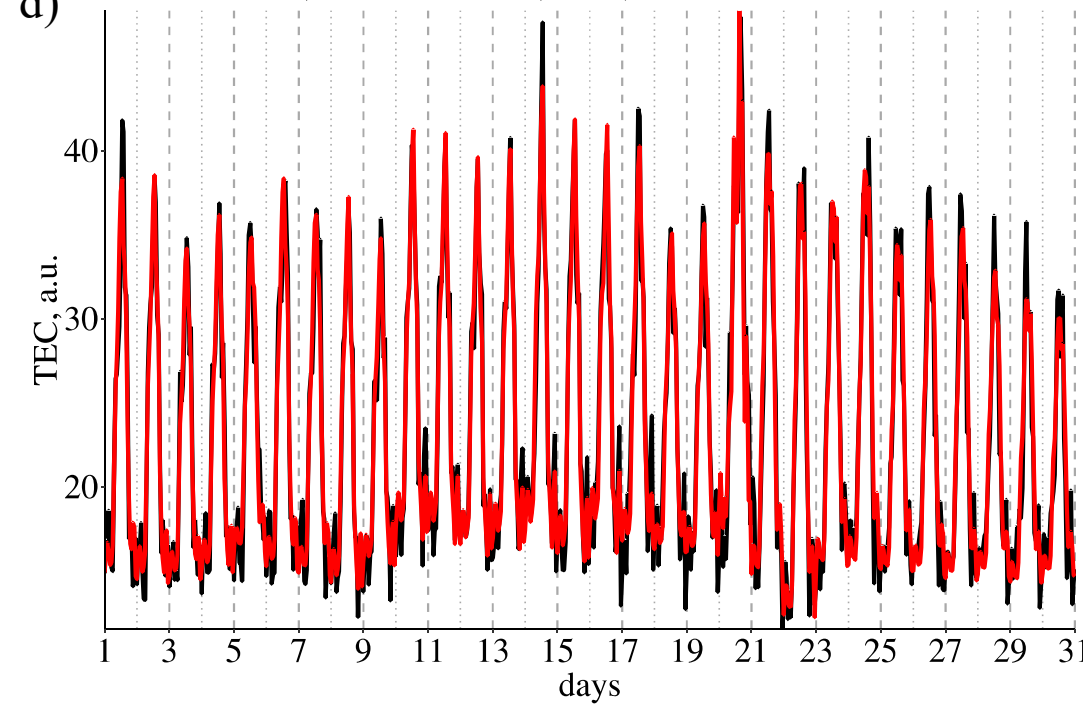

**Figure3.**

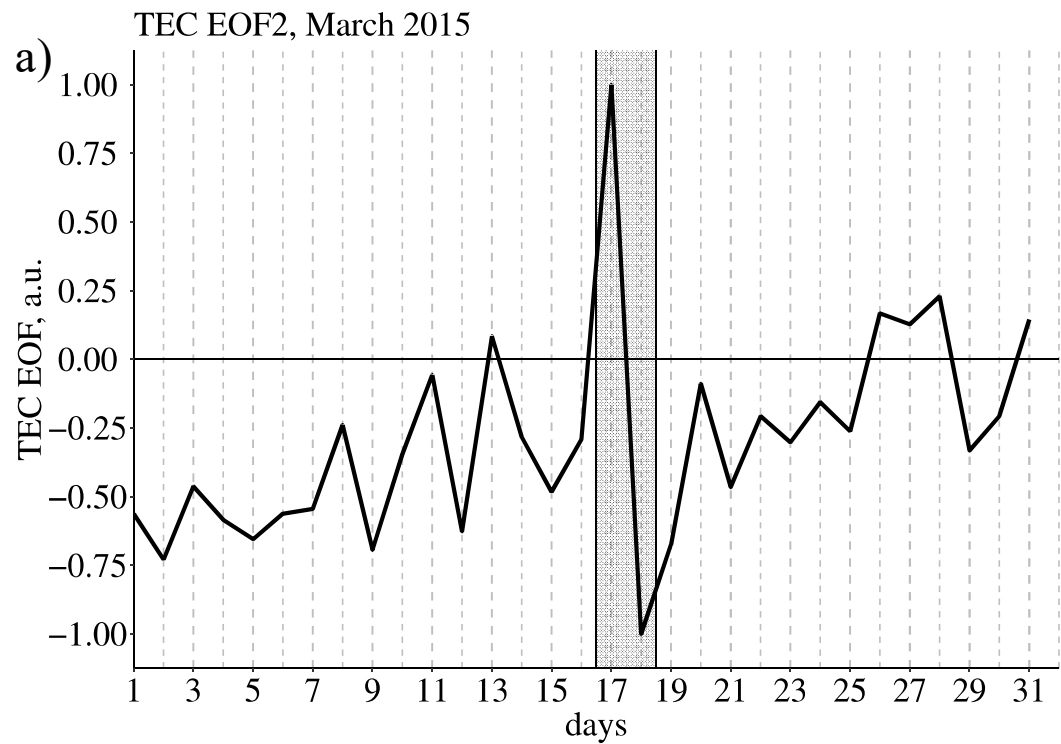

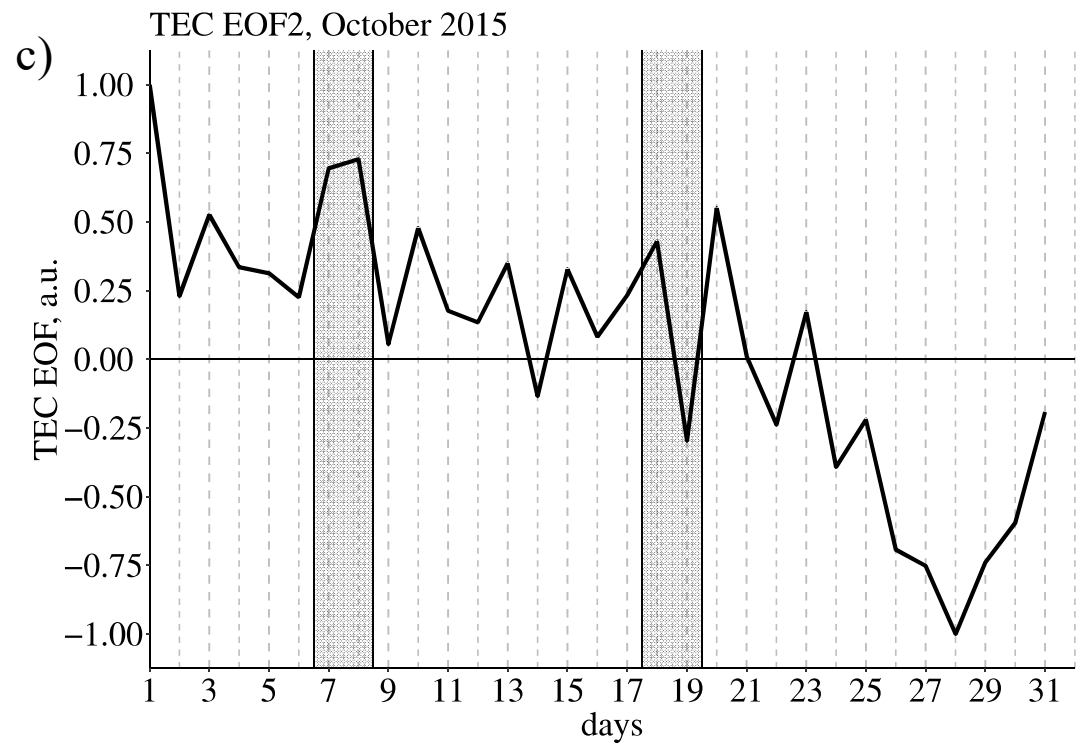

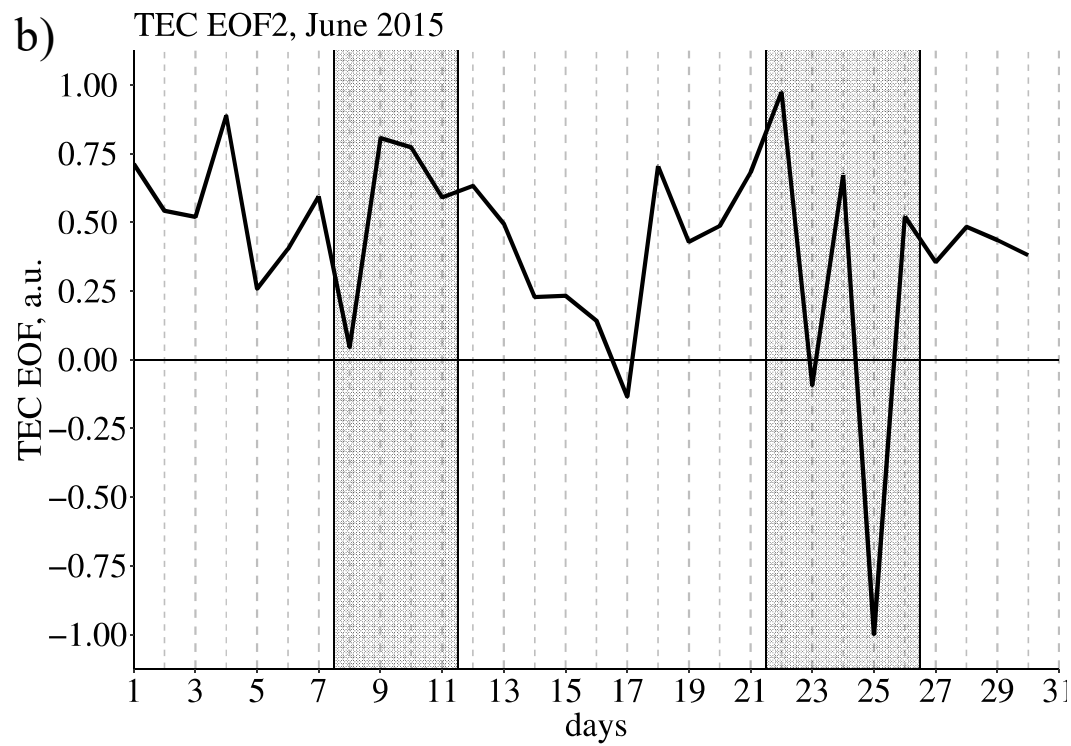

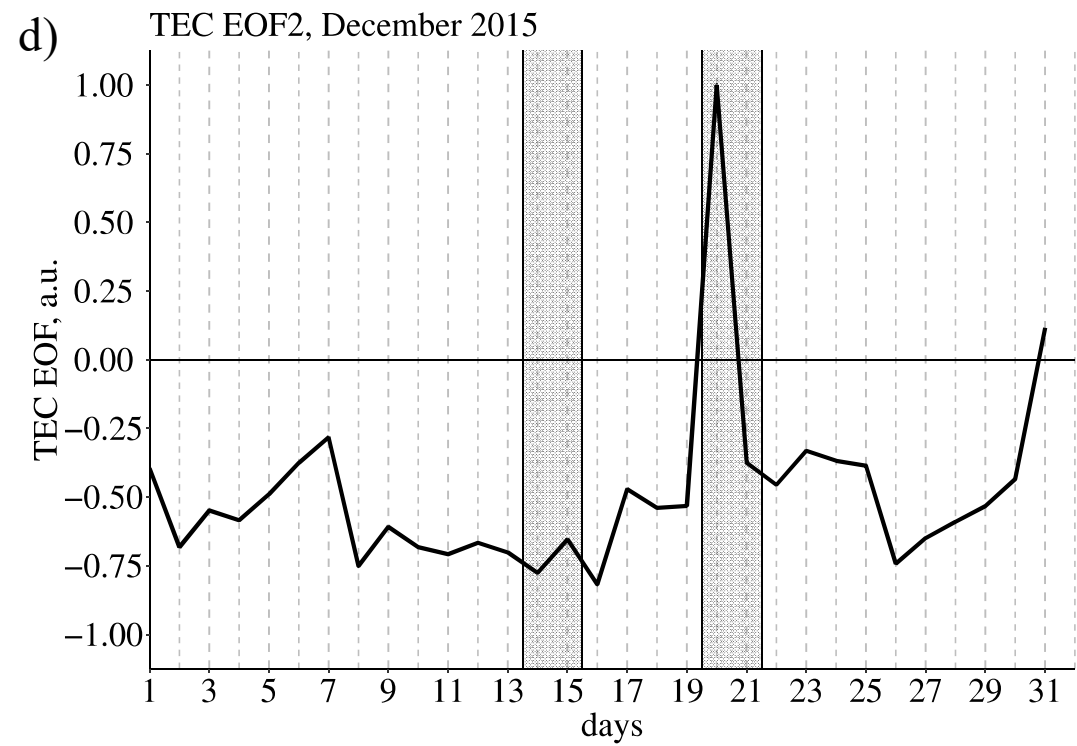

**Figure2.**

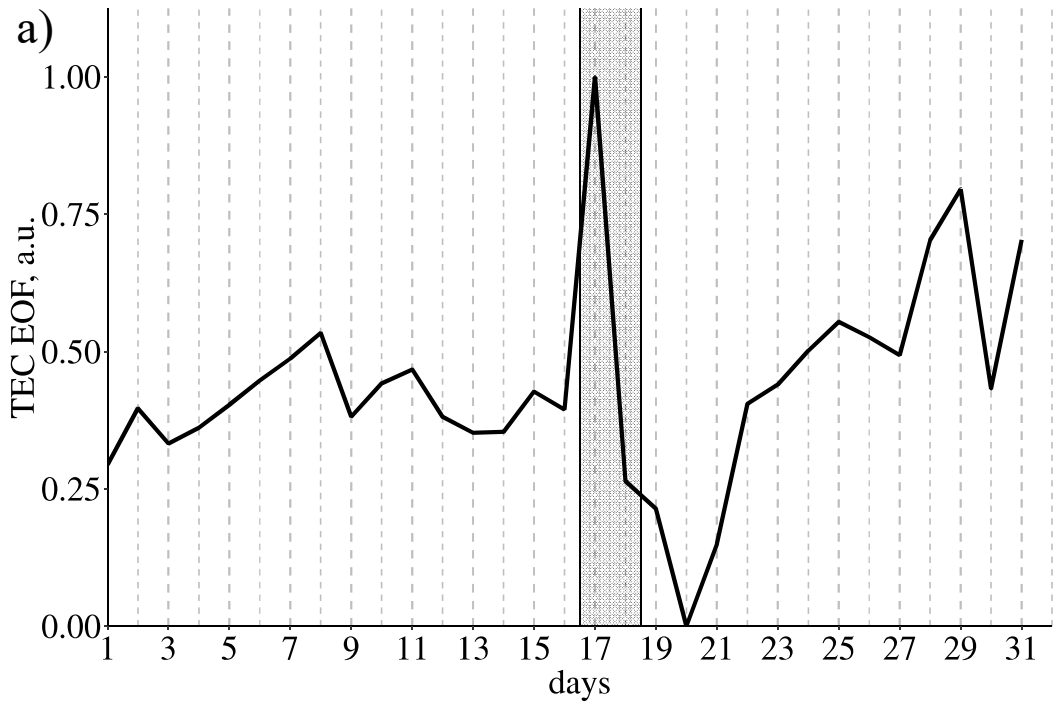

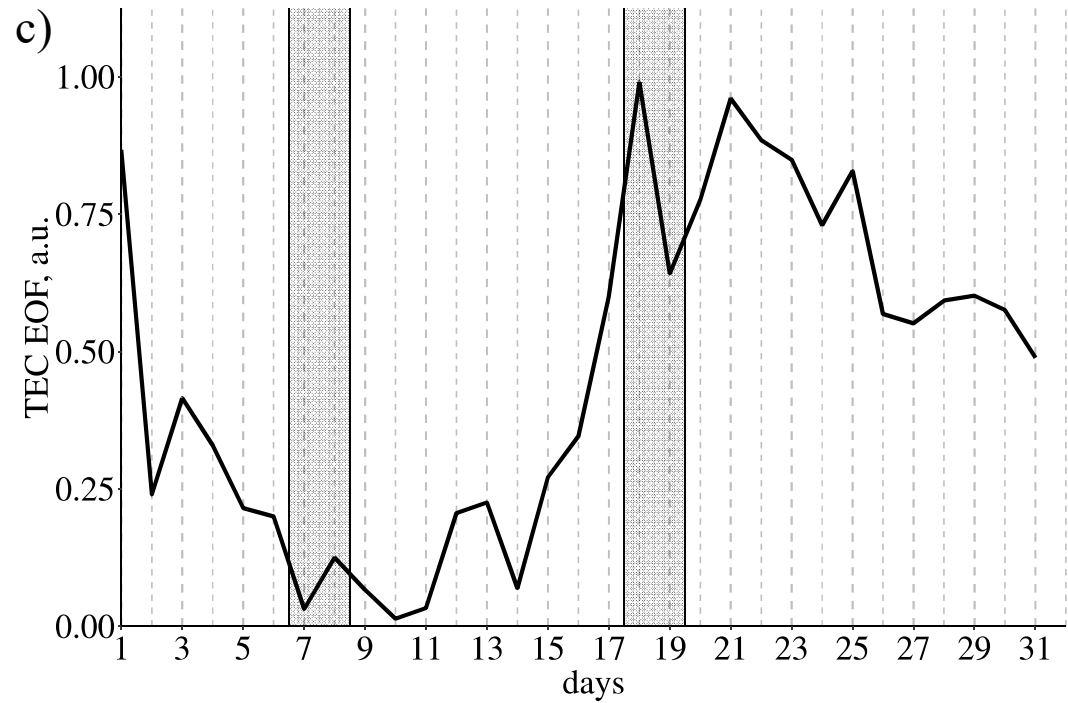

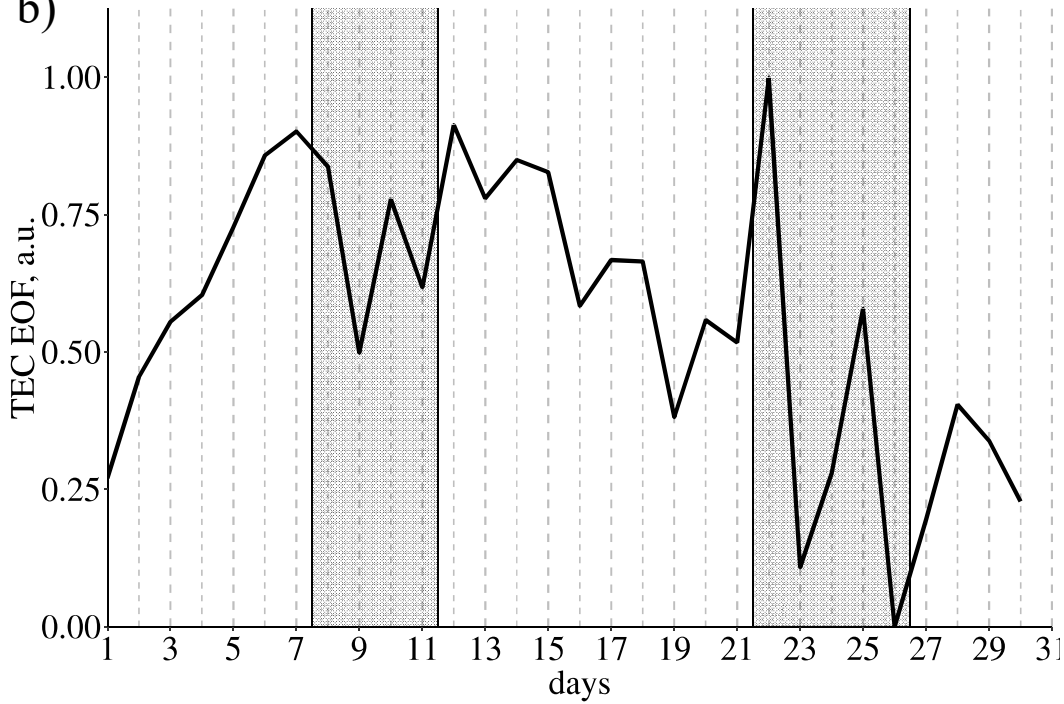

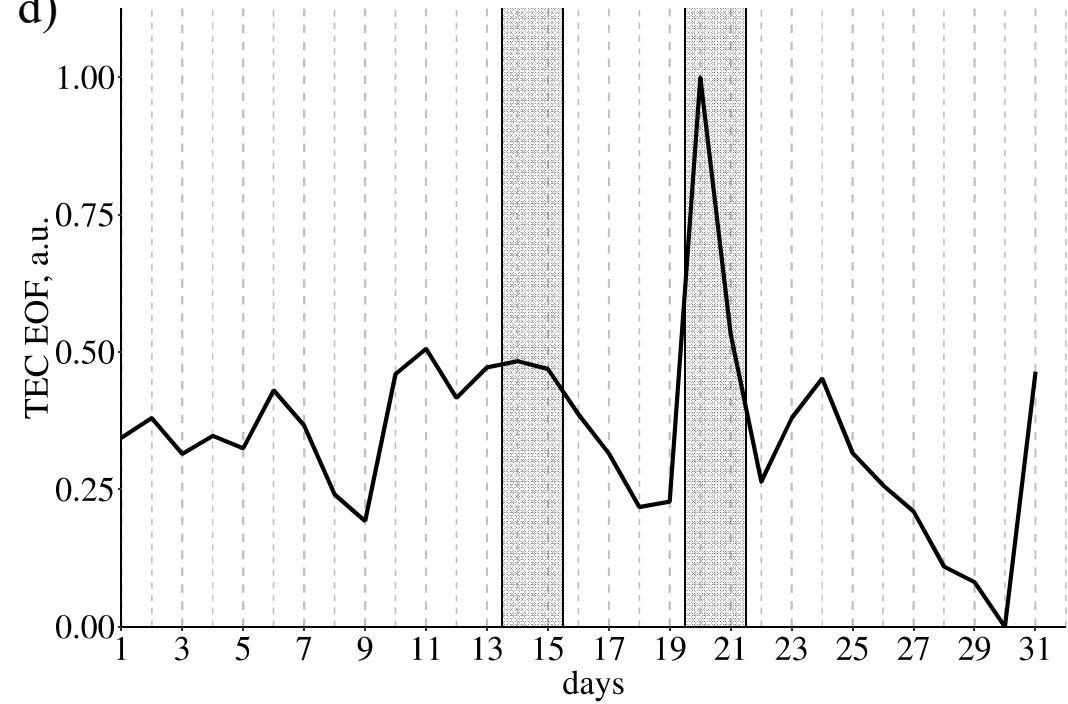

**Figure1.**

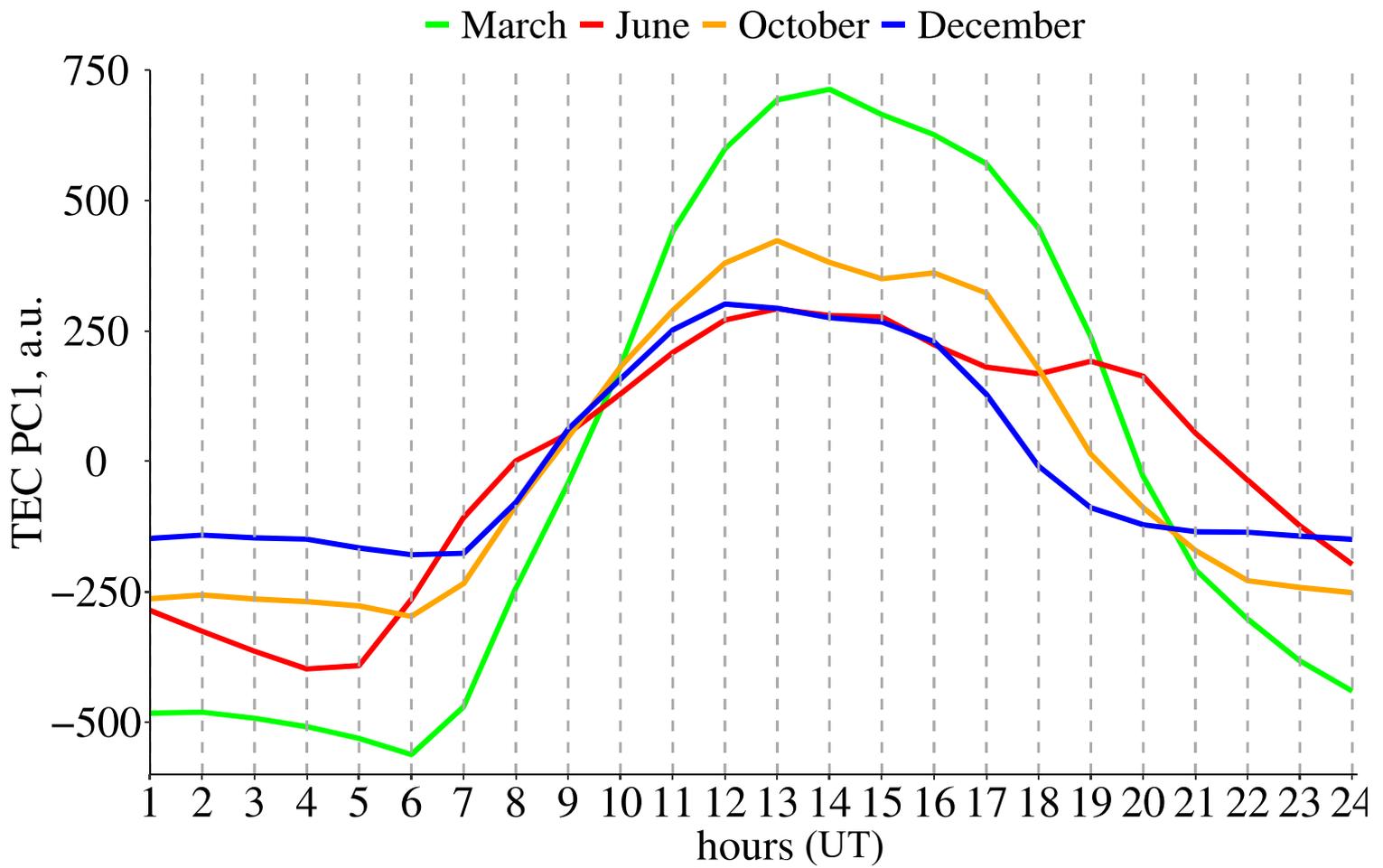

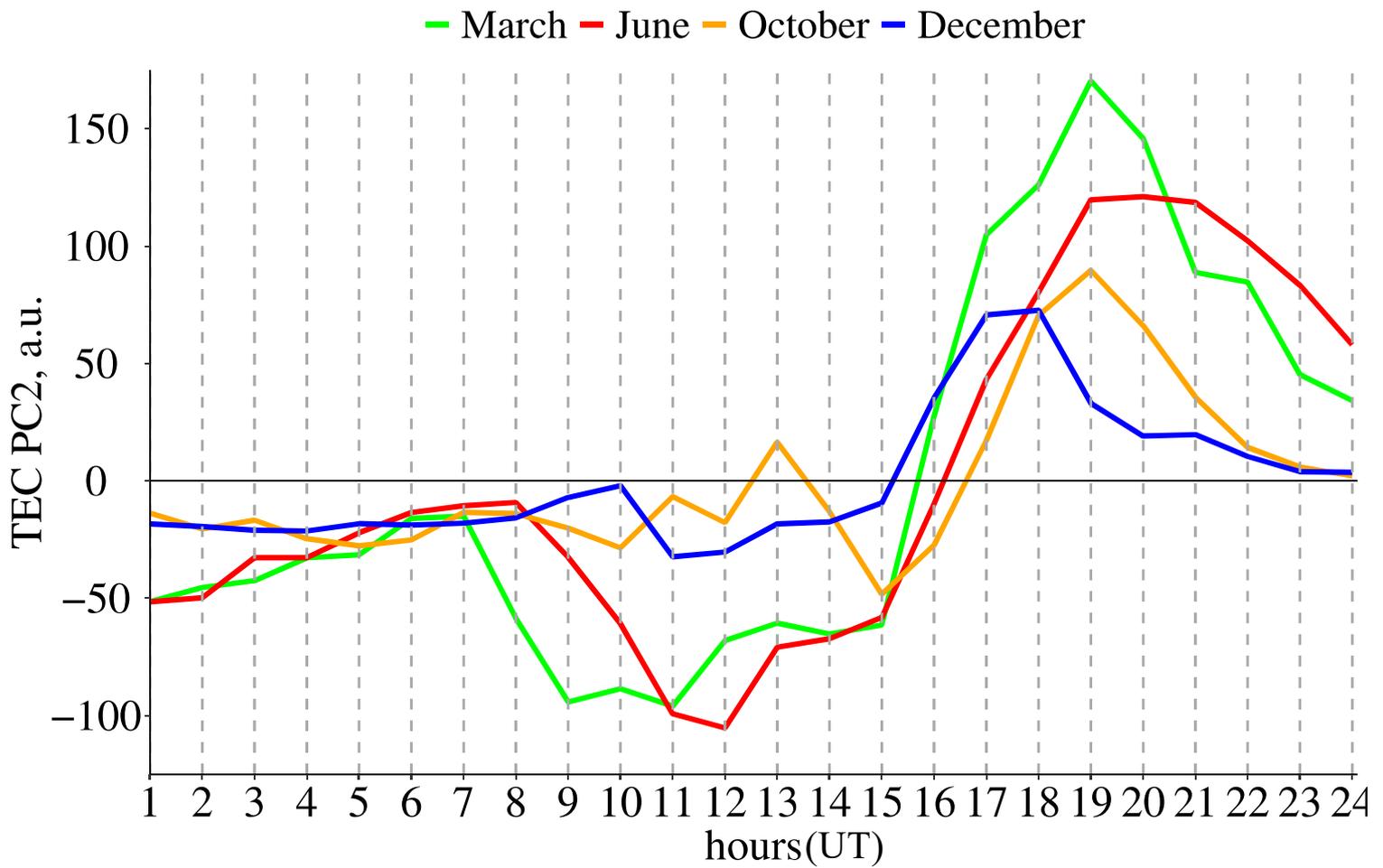